\documentclass[10pt,journal,compsoc]{IEEEtran}

\usepackage{multirow}
\usepackage{amsmath,amssymb}

\usepackage{algpseudocode}
\usepackage[ruled,linesnumbered]{algorithm2e}
\usepackage[justification=centering]{caption}
\usepackage{url} 
\usepackage[section]{placeins}
\usepackage{longtable}
\usepackage{tablefootnote}
\usepackage{booktabs}
\usepackage{tabularx}

\ifCLASSOPTIONcompsoc
  \usepackage[nocompress]{cite}
\else
  \usepackage{cite}
\fi

\ifCLASSINFOpdf
   \usepackage[pdftex]{graphicx}
   
\else
\fi

\usepackage{array}

  \usepackage[caption=false,font=footnotesize,labelfont=sf,textfont=sf]{subfig}
  \usepackage[caption=false,font=footnotesize]{subfig}

\newcommand\MYhyperrefoptions{bookmarks=true,bookmarksnumbered=true,
	pdfpagemode={UseOutlines},plainpages=false,pdfpagelabels=true,
	colorlinks=true,linkcolor={blue},citecolor={blue},urlcolor={blue},
	pdftitle={Big Data Meets Metaverse: Survey, Security, and Opportunities},
	pdfsubject={Typesetting},
	pdfauthor={Jiayi Sun},
	pdfkeywords={Metaverse, big data, survey, security and privacy, opportunities.}}
\ifCLASSINFOpdf
\usepackage[\MYhyperrefoptions,pdftex]{hyperref}
\else
\usepackage[\MYhyperrefoptions,breaklinks=true,dvips]{hyperref}
\usepackage{breakurl}
\fi

\usepackage{url}

\begin{document}
\title{Big Data Meets Metaverse: A Survey}

\author{Jiayi Sun, Wensheng Gan*, Zefeng Chen, Junhui Li, Philip S. Yu,~\IEEEmembership{Fellow,~IEEE}
	
	\IEEEcompsocitemizethanks{
		
	\IEEEcompsocthanksitem Jiayi Sun, Zefeng Chen, and Junhui Li are with the College of Cyber Security, Jinan University, Guangzhou 510632, China. 
	
	\IEEEcompsocthanksitem Wensheng Gan is with the College of Cyber Security, Jinan University, Guangzhou 510632, China; and with Pazhou Lab, Guangzhou 510330, China.  (E-mail: wsgan001@gmail.com) 

	\IEEEcompsocthanksitem Philip S. Yu is with the University of Illinois at Chicago, Chicago, USA.} 
	
	\thanks{Corresponding author: Wensheng Gan}
}

\IEEEtitleabstractindextext{%
\begin{abstract}
	
We are living in the era of big data. The Metaverse is an emerging technology in the future, and it has a combination of big data, AI (artificial intelligence), VR (Virtual Reality), AR (Augmented Reality), MR (mixed reality), and other technologies that will diminish the difference between online and real-life interaction. It has the goal of becoming a platform where we can work, go shopping, play around, and socialize. Each user who enters the Metaverse interacts with the virtual world in a data way. With the development and application of the Metaverse, the data will continue to grow, thus forming a big data network, which will bring huge data processing pressure to the digital world. Therefore, big data processing technology is one of the key technologies to implement the Metaverse. In this survey, we provide a comprehensive review of how Metaverse is changing big data. Moreover, we discuss the key security and privacy of Metaverse big data in detail. Finally, we summarize the open problems and opportunities of Metaverse, as well as the future of Metaverse with big data. We hope that this survey will provide researchers with the research direction and prospects of applying big data in the Metaverse.

\end{abstract}

}

\maketitle

\IEEEdisplaynontitleabstractindextext

\IEEEpeerreviewmaketitle

\section{Introduction}

\IEEEPARstart{I}{n} 1992, Neal Stephenson coined the term ``Metaverse" in the science fiction novel ``Snow Crash" \cite{stephenson2003snow}. In the novel, humans appear as virtual avatars, communicating and interacting in a three-dimensional virtual space with the real world as their background. But digital technology was not enough to support this vision, and the theory failed to materialize. In 2003, Linden Labs developed a virtual game called Second Life\footnote{https://secondlife.com/}, which implemented a virtual world with social, entertainment, and production functions. Players participate in the game as digital avatars. In recent years, technologies such as the Internet, artificial intelligence (AI) \cite{zhang2021study,huynh2022artificial}, extended reality (XR) \cite{ratcliffe2021extended,xi2022challenges} and blockchain (BC) \cite{mohanta2019blockchain, gadekallu2022blockchain} have developed rapidly. Various technologies merge and influence each other, providing the basis for the realization of the Metaverse. At the same time, due to the COVID-19 pandemic, the construction of virtual scenes has been given more attention. 2021 is known as the first year of the Metaverse. This year, the Metaverse has attracted wide attention, and related applications have emerged one after another. In March 2021, Roblox\footnote{https://www.roblox.com/} went public on Nasdaq in the United States in what has been described as the ``First Metaverse stock". In October 2021, Facebook changed its name to Meta, which was taken from the prefix of the Metaverse. The concept of the Metaverse appears in all walks of life.

Simply put, the Metaverse is a shared, immersive, online virtual world facilitated by the convergence between Internet and Web technologies, as well as XR \cite{lee2021all}. The Metaverse is the inevitable product of 5G, AI, edge computing \cite{du2018big}, and other technologies when they reach a certain stage of development. It is a complex virtual world and a digital world that are parallel and mutually mapped with the real world under the accumulation of emerging technologies. It is also an important technical basis for the development of massive data in the future. In the virtual world of the Metaverse, it is far more digital than in the real world. The spatial structures, scenes, and characters outlined by digital technology are essentially in the form of data \cite{gadekallu2022blockchain}. At a technical level, the Metaverse can be seen as a fusion carrier of big data and information technology. The user's information and movements in the Metaverse are recorded in files in the form of data. With the increase of users, the metaverse will generate massive amounts of data, thus forming a big data network. This network will continue to grow, posing a huge challenge to data processing technology \cite{ooi2022sense}. The seamless connection between the virtual and the real in the Metaverse requires the support of numerous IoT devices, which collect and process data in the physical world in real time. The integration and application of offline and online data is a key task of big data technology. So the ability to process big data is very important to the Metaverse \cite{mohammadi2018deep}. In addition, with the increase of data, people need to use intelligent data analysis tools to get useful information, make decisions more predictable and more accurate, and more effectively guide all aspects of production and life. As a result, big data technology is one of the key technologies for the successful implementation of the Metaverse \cite {han2022dynamic, cai2022compute}.

``Big data" refers to those data sets that are particularly large in volume and complex in data categories. It can also be described as a massive, high-growth and diversified information asset \cite{sagiroglu2013Big}. This kind of data set can not be acquired, managed, and processed by traditional data analysis tools. It needs to apply a new processing mode to the data in order to get more decision-making information. Big data contains a large amount of unstructured, semi-structured, or structured data from a variety of sources. Big data analytics transforms big data into intelligent data to find hidden information that can be used in machine learning projects and to build predictive modeling \cite{hariri2019uncertainty}. According to the actual operation process, big data analysis can be divided into four parts: storage, cleaning, analysis, and visualization \cite{ge2018big}. Big data storage technology solves the problems of data access speed and capacity. Due to the diversity and accuracy of data, data cleaning provides solutions for different data sources and data quality problems. Data analytics and visualization aim to extract deeper values from big data and provide accurate and predictive guidance for human activities.

Big data will inject vitality into the development of digital space and promote the convergence of many fields. As the core of big data applications, big data analytics is characterized by the analysis of massive and complete datasets to obtain the key information from them \cite{katal2013big}. The Metaverse is a digital world parallel to the real world, made entirely of data. The large scale of data sources and isolated information in the Metaverse will bring specific difficulties, which put forward high requirements for real-time data collection, processing, and analysis \cite{ning2021survey}. The application of big data can solve the problems faced by massive data in the Metaverse. Big data infrastructure has options to analyze and process a wide variety of data types, while mining information and pooling experience to get the most from Metaverse and assist in predicting future outcomes \cite{hajjaji2021big}. Therefore, big data is one of the key technologies indispensable to the successful application of the Metaverse. In addition, data security has always been a hot issue in both the Metaverse and big data \cite{gadekallu2022blockchain}. Inevitably, the combination of big data and the Metaverse should also focus on security solutions.

\textbf{Research status:} In this work, we survey existing papers on the Metaverse and data. At the technical level, there have been several papers focusing on the data part of the Metaverse. Ooi \textit{et al.} \cite{ooi2022sense} summarize the relevant theories and technologies that facilitate the connectivity between the Metaverse and the real world from the perspective of data. Han \textit{et al.} \cite{han2022dynamic} propose a dynamic resource allocation framework to accomplish the synchronization between services and data of the Metaverse and IoT. Cai \textit{et al.} \cite{cai2022compute} outline a cloud network flow mathematical framework that provides a solution to the compute-and data-intensive network of the Metaverse. Sun \textit{et al.} \cite{sun2022matrix} propose a multichain data aggregator. It supports blockchain-based Metaverse that bridge the data access gap between blockchain and end applications. Gadekallu \textit{et al.} \cite{gadekallu2022blockchain} describe the application of blockchain to data processing in the Metaverse. At the application level, there are some papers that apply the Metaverse and data science. Park \textit{et al.} \cite{park2022method} study ways to create Metaverse using smartphone data. Yang \textit{et al.} \cite{yang2022smart} implement a smart medical system based on Metaverse, data science, and AI. Angelini \textit{et al.} \cite{angelini2022towards} propose a framework for visualizing, recording, and synchronizing data from experiences and human signals in VR. In addition, we surveyed existing reviews, as shown in Fig. \ref{table:Gaps}. Gadekallu \textit{et al.} \cite{gadekallu2022blockchain} introduce the data flow of the Metaverse and the application of big data technology in the Metaverse from the perspective of blockchain. However, there is no review to give the technical framework of the Metaverse from the perspective of data and introduce the technical point of big data in the Metaverse.

\textbf{Contributions:}  To the best of our knowledge, this is the first review examining the combination of the Metaverse and big data. Our main purpose is to investigate the feasibility of combining big data and the Metaverse and the promising solutions and new insights that this combination brings to the development of the Metaverse, explore hidden opportunities and challenges, and identify future research directions. This survey aims to help researchers and practitioners better understand and implement big data concepts in the Metaverse. The contributions of our work are summarized as follows:

\begin{itemize}
	\item Although both of them have been widely studied, fewer studies have joined the Metaverse with big data. A review article related to both hot topics has not been proposed yet. To the best of our knowledge, we are the first to survey recently published major research on data science and the Metaverse.
	
	\item We summarize the main roles and key requirements of big data in the Metaverse. We show that big data is ubiquitous, and the Metaverse is the future technology development trend.
	
	\item We discuss the integration of other supporting technologies of the Metaverse and big data. Moreover, several key security and privacy issues of Metaverse big data are summarized in detail.
	
	\item We highlight the open view and future research direction of the combination of big data and Metaverse.
\end{itemize}

\textbf{Roadmap:} Section \ref{sec:bigdata} introduces the basic concepts, benefits, and common platforms of big data. Section \ref{sec:Metaverse} briefly reviews what the Metaverse is, its main benefits, and its current applications. The relationship between big data and the Metaverse and what it takes for the Metaverse to meet big data are respectively reviewed in Section \ref{sec:Makesense}. We mention the security and privacy issues when dealing with big data in the Metaverse in Section \ref{sec:Security}. In Section \ref{sec:Opportunities}, we discuss the open problems and opportunities of   big data meets Metaverse in detail. Conclusions are provided in Section \ref{sec:conclusion}.

\section{Big Data} \label{sec:bigdata}

\subsection{What is Big Data?}

Big data is a concept that deals with formatting, storing, and analyzing large datasets \cite{el2018investigating}. The definition of big data refers to a collection of data which is so large that it cannot be acquired, managed, and processed with the help of conventional software within a certain period of time. It was first used in academia when Bryson \textit{et al.} \cite{bryson1999visually} explored gigabyte data sets visually in real time during an ACM communication in 1999. And it is well known that Doug Laney \cite{russom2011big} first proposed the three V's, including volume, velocity, and variety, in 2011.

\begin{table*}
	\centering
	\caption{Contributions and gaps of existing surveys}
	\begin{tabular}{|c|c|p{5cm}|c|c|c|c|}
		\hline
		\textbf{Ref.} & \textbf{Year}   & \textbf{One-sentence summary}  & \textbf{Technology} & \textbf{Applications} & \textbf{Security and privacy} & \textbf{Big data} \\ \hline
		
		\cite{lee2021all}& 2021  & A complete survey on technological singularity, virtual ecosystem, and research agenda  & \checkmark  & $\times$ &  \checkmark & $\times$       \\ \hline		
		
		\cite{ning2021survey}   & 2021  & A Survey on Metaverse: the state-of-the-art, technologies, applications, and challenges  & \checkmark   & \checkmark &  $\times$ & $\times$         \\ \hline
		
		\cite{park2022Metaverse}& 2022   &  A Metaverse: Taxonomy, components, applications, and open challenges  & \checkmark   & \checkmark &  $\times$  & $\times$       \\ \hline		
		
		\cite{wang2022survey}& 2022   & A survey on Metaverse: Fundamentals, security, and privacy & \checkmark   & \checkmark &  \checkmark & $\times$      \\ \hline
		
		\cite{di2021Metaverse}& 2021 & Metaverse: Security and privacy issues & \checkmark   & $\times$ &  \checkmark & $\times$      \\ \hline
		
		\cite{zhao2022Metaverse}  & 2022    &  Metaverse: Security and Privacy Concerns    & $\times$   & $\times$  & \checkmark & $\times$     \\ \hline		
		\cite{huynh2022artificial}    & 2022   &  Artificial intelligence for the Metaverse: A survey   & \checkmark   & \checkmark &  $\times$ & $\times$         \\ \hline
		
		\cite{gadekallu2022blockchain}   & 2022   &  Blockchain for the Metaverse: A review   & \checkmark  & $\times$ &  \checkmark  & \checkmark     \\ \hline
		
		\cite{yang2022fusing} & 2022   & Fusing blockchain and AI with Metaverse: A survey & \checkmark   & $\times$ & $\times$ & $\times$        \\ \hline
		
		\cite{jeon2022blockchain}    & 2022    & Blockchain and AI meet in Metaverse   & \checkmark   & $\times$ &  \checkmark  & $\times$    \\ \hline
		
		Our work & 2022  & Big data meets Metaverse: A survey &\checkmark & \checkmark &  \checkmark & \checkmark\\ \hline
	\end{tabular}
	
	\label{table:Gaps}
\end{table*}

In terms of size, the volume or scale of data has exceeded terabytes and petabytes. Hence, because of the large scale, only high velocity can keep up with the transmission of large data. In addition, the large volume also results in traditional data processing software just being unable to manage them. However, as big data technology has become more stable and mature in recent years, these massive volumes of data can be used to cope with problems that have never been solved by all walks of life. The variety of big data is reflected in the diversity of sources and forms. With the rapid development of the Internet, data is everywhere in life and work, such as various transaction records, e-mails, music, pictures, videos, and application software records, etc. In a data form, big data can be broadly divided into the following three data structures \cite{al2019big, younas2019research}.

Big data combines three kinds of data, which are structured, semi-structured, and unstructured data, respectively. This data is collected to mine the information and used for applications for advanced analytics as well as projects of machine learning and predictive modeling, etc. \cite{al2019big, mhammedi2021heterogeneous}. Generally speaking, structured data can be represented and stored by relational databases. This type of data can be expressed by two-dimensional tables logically, such as transaction and financial records, etc. Unstructured data, a data type without a fixed structure, is another form of data representation. For instance, office documents, text, various images, and audio or video information are all unstructured data. And as the compromised data type between completely structured data and completely unstructured data, semi-structured data does not represent data in the form of relational databases or other data tables, but it contains related labels. Therefore, it is also called a self-describing structure \cite{rusu2013converting, kumar2021integrated}.

\subsection{Big Data Benefits}

The sources of big data are very broad. Data is from various applications, such as social platforms, e-mails, trading platforms, sensors, audio, video, images, etc. This data is stored in a rapidly growing database \cite{agrawal2011challenges}. Through the capture, formation, storage, management, analysis, sharing, and visualization of big data, it helps to obtain more useful information \cite{oguntimilehin2014review}. Organizations in any industry with big data can benefit from this information to gain insight and depth to solve real-world problems \cite{choi2018big, grover2018creating, sarker2021data}.

Big data technology has appeared in all aspects of people's lives and is applied in various industries. Big data is mainly used in daily production \cite{zhong2016visualization}, economy \cite{nobre2017scientific}, Internet industry \cite{aceto2020industry}, e-commerce \cite{zheng2020commerce}, and other activities \cite{jiang2018rethinking}. Through the collection, processing, and integration of massive data, it combines with artificial intelligence, cloud computing, and other technologies to conduct an in-depth analysis of the data \cite{yanhua2020application, cao2021innovation}.

\begin{figure}[t]
	\centering
	\includegraphics[clip,scale=0.26]{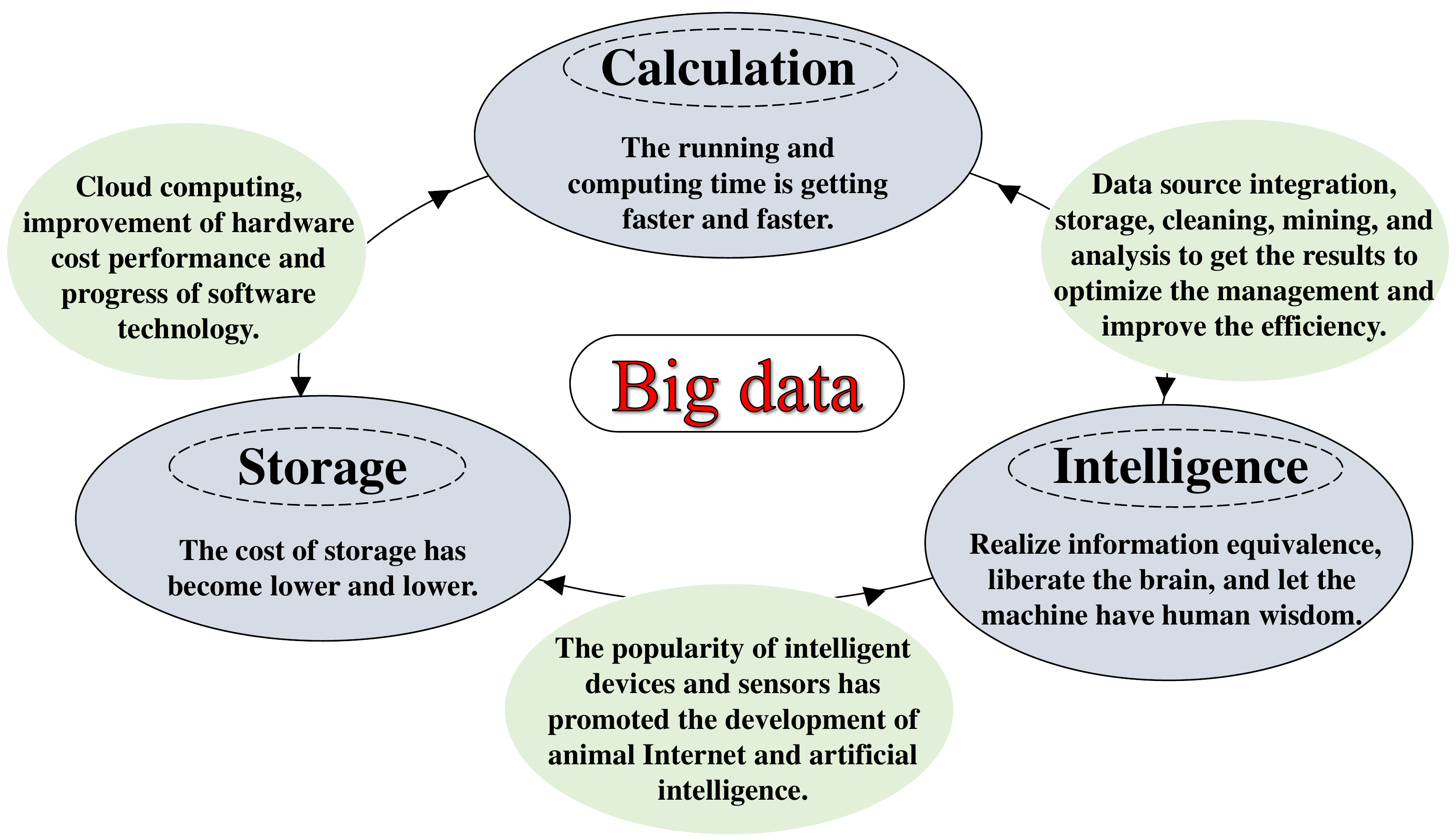}
	\caption{The operation of big data}
	\label{Bigdata}
\end{figure}

The significance of big data lies not only in mastering massive data information but also in how to accurately and professionally process these datasets. After a large amount of data is collected, it is necessary to process and support the data through storage, computing, and artificial intelligence to realize the value-added of the data. In a digital world, the essence of the Metaverse is to constantly generate and process data. Big data technology is required to fully utilize these data in order to realize value. These technologies include computing power, storage power, and intelligence power. At present, as Figure \ref{Bigdata} shows, big data is actually a process of continuous development. The decline in storage costs has led to a higher demand for cloud computing and software technology, which also promotes computing speed. The increase in computing speed requires the emancipation of the brain, which makes the machine more and more intelligent. The promotion of intelligence has further promoted the development of storage capacity in terms of the popularity of intelligent devices, sensors, and other hardware. Such reciprocity has made big data technology develop continuously.

In recent years, the concepts of the Metaverse and Web 3.0 have taken the world by storm \cite{kim2021advertising, parthasarathy2022web}. The Metaverse is the next-generation Internet, and Web 3.0 is the spiritual core of the next-generation Internet \cite{kim2021advertising, cook2020spatial}. Although the Metaverse and Web 3.0 are grand, they are not empty talk. They need a lot of current technology accumulation, such as the visual presentation of data on the chain, natural language processing for virtual human interaction, data security, etc., all of which are inseparable from big data.

\subsection{Big Data Platform}

With the development of the big data industry, the related technologies of the big data ecosystem have been making iterative progress \cite{singh2015survey}. Its core technologies can be divided into the nine categories in Table \ref{platform}. 

\section{Metaverse} \label{sec:Metaverse}
\subsection{What is the Metaverse?}

The term Metaverse is made up of two words: Meta and Verse, where Verse is short for universe\footnote{https://en.wikipedia.org/wiki/Metaverse}. Its goal is to create a digital space parallel to the real world. This virtual space has a complete social system, such as an economic system, identity system, and so on, and can interact with the real world \cite{wang2022metasocieties}. As the next stage of the Internet, the concept of the Metaverse is revolutionary. When completed, it is expected to be a virtual place where people can work, play, socialize, study, and shop. The Metaverse is a mixture of computer-generated reality, extended reality, and physical reality that will blur the lines between networks and reality.

In Chinese, the word ``meta" means first, beginning, important, and consummation. On the one hand, it represents a new beginning; on the other hand, from the perspective of completeness, its connotation already includes not only the virtual world and the past world, but also the real world and the future world. At the technical level, it includes many emerging technologies such as virtual reality, augmented reality, mixed reality, blockchain, cloud computing, digital twin, artificial intelligence, and so on \cite{lee2021all,ning2021survey}. In short, it is a large integration of human, virtual, and reality across time and space.

\begin{figure}[h]
	\centering
	\includegraphics[scale = 0.36]{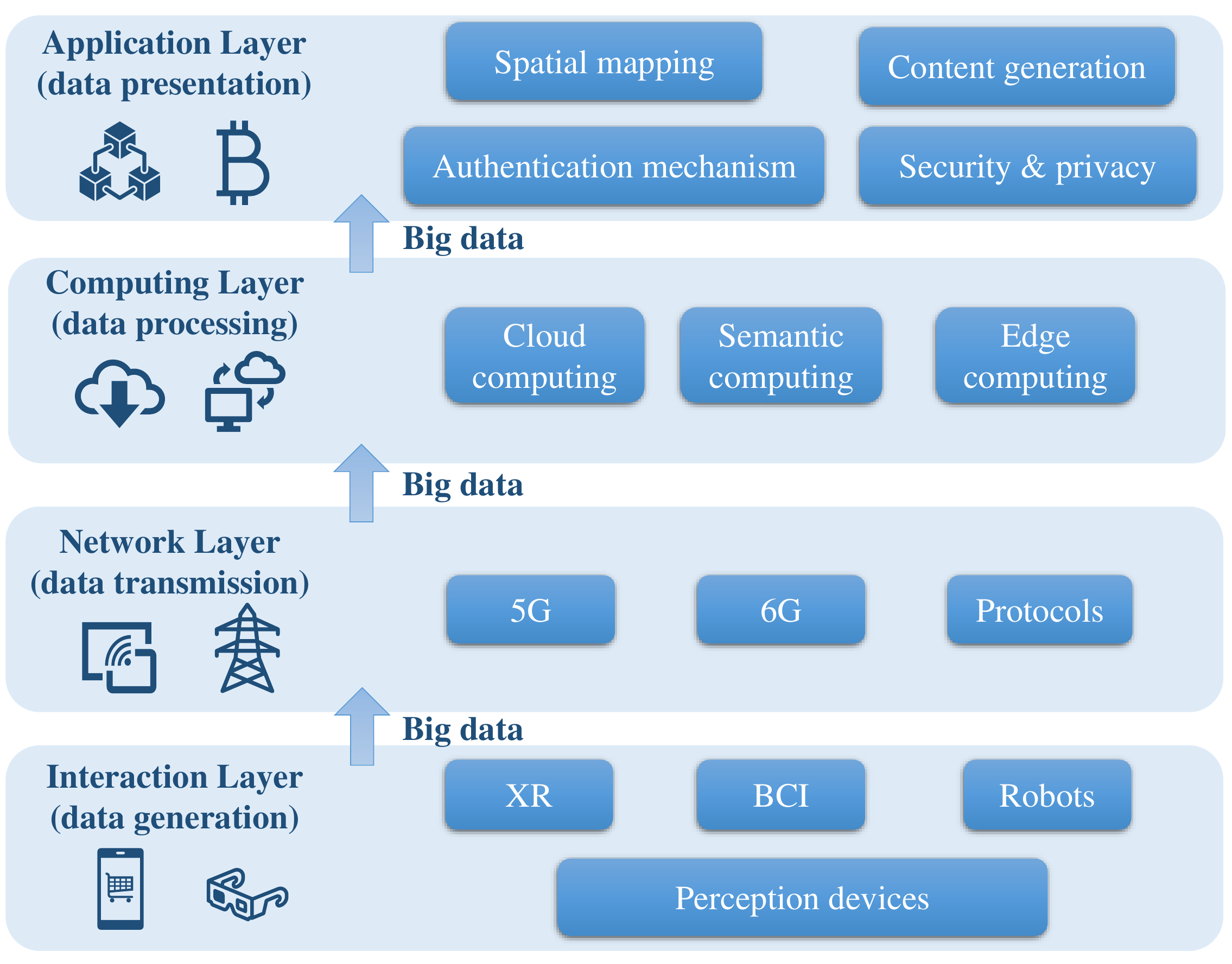}
	\caption{The framework of the Metaverse}
	\label{fig:Framework}
\end{figure}

From a technical perspective, the Metaverse can be divided into four layers: interaction layer, network layer, computing layer, and application layer, as shown in Fig. \ref{fig:Framework}. The interaction layer enables interaction between the physical and digital worlds. Users can manipulate avatars using brain-computer interfaces, XR, robots, and other devices to move around virtual worlds and create digital footprints. Sensing devices in the IoT can also capture data about the user's behavior in the physical world and upload it to the Metaverse. The network layer is the guarantee of the Metaverse's real-time. The Metaverse requires a reliable and high-speed network environment that seamlessly connects the physical and the virtual and between different virtual worlds. In the computing layer, cloud computing, edge computing, and artificial intelligence are used to process and analyze numerous heterogeneous data sources to achieve data interoperability and obtain useful information from them. Data presentation is done in the application layer, such as spatial mapping, content production, and authentication mechanisms. Data is the key to the interaction between reality and virtuality, and the generation, transmission, processing, and presentation of data correspond to the above four layers respectively.

After widespread application of the Metaverse, rich information about users will be collected in the Metaverse, which is the most valuable part of the Metaverse. Companies can use this information to personalize services and create unique experiences for users. Data science and artificial intelligence will play an important role in extracting useful information from massive amounts of data \cite{huynh2022artificial,yang2022smart}. In addition, the cloud data platform is the basis of data management and analysis to meet the needs of collecting, storing, and analyzing numerous heterogeneous data in the Metaverse. This can break down an isolated data island and gain meaningful insights that can help predict the future.

\begin{table*}[h]
	\caption{Framework of Big Data Platform}
	\label{platform}%
	\begin{tabularx}{\textwidth}{m{2.8cm}<{\centering}m{6.8cm}<{\raggedright}m{6.2cm}<{\raggedright}}
		\toprule
		\hline
		\textbf{Framework} & \multicolumn{1}{c}{\textbf{Effect}} & \multicolumn{1}{c}{\textbf{Example}}  \\
		\hline
		\small{\textcircled{\small{\footnotesize{1}}}} \footnotesize{Data collection} & Data acquisition is also known as data synchronization. After generating massive data, the first step is to collect the data, so data collection is the basis of big data.  	& The Internet \cite{qi2020overview}, mobile applications \cite{salo2018users} and geolocation \cite{luceri2018vivo} are the most common methods of collecting massive amounts of data \cite{ribeiro2021towards}.	\\ 	\hline
		
		\small{\textcircled{\small{\footnotesize{2}}}} \footnotesize{Data storage} & The rapid growth of data has brought great challenges to data storage, which has led to the emergence of high-storage and distributed storage systems. & At present, with the popularization of cloud storage technology, mobile cloud storage solves a large part of the security data sharing and data storage problems \cite{feng2022secure}.  \\ \hline
		
		\small{\textcircled{\small{\footnotesize{3}}}} \footnotesize{Distributed resource management}  & In the traditional IT field, the server resources of enterprises are limited and fixed. However, the application scenarios of the server are flexible and changeable. The demand for temporary tasks has increased greatly, and these tasks often require large amounts of server resources. Therefore, distributed resource management systems and frameworks have emerged.
		& Recently, optimized task scheduling for distributed resource management in cloud \cite{mijuskovic2021resource} and fog-assisted \cite{wadhwa2022optimized} has proposed to improve the quality of service, which can take into account the dynamic characteristics of tasks to achieve scheduling in real-time scenarios.\\    \hline
		
		\small{\textcircled{\small{\footnotesize{4}}}} \footnotesize{Data computing}  & After big data is collected and stored, it needs to be calculated. This large amount of calculation requires high computing power, which requires the support of the data computing technology framework. As a result, many upgrades and evolutions of data computing technologies have also been born.
		& Data computing includes offline and real-time data computing. Spark is mainly used for offline computing in enterprises, and Flink is mainly used for real-time computing \cite{chintapalli2016benchmarking}. In addition, the evolution of emerging computing paradigm cloud to fog has been studied \cite{kumar2021evolution}. \\  \hline
		
		\small{\textcircled{\small{\footnotesize{5}}}} \footnotesize{Data analysis} & These techniques facilitate the effective application of data. Through the analysis of business figures and data, problems can be found in time and solutions can be sought. The data is transformed into quantitative indicators for evaluation, which plays a key role in decision-making. & Data analysis is widely used in IoT \cite{xie2022research}, smart city \cite{li2022big}, blockchain \cite{balcerzak2022blockchain}, intelligent manufacturing \cite{chen2022influence}, etc. Different data technology frameworks include Hive, Impala, Kylin, Clickhouse, Druid, Doris, etc. They have different application scenarios for offline and real-time data analysis. \\ \hline
		
		\small{\textcircled{\small{\footnotesize{6}}}} \footnotesize{Task Scheduling} & The task scheduling technology framework is suitable for routine tasks executed at regular intervals, as well as multi-level tasks containing complex dependencies for scheduling.  & Task scheduling in cloud computing \cite{ibrahim2021task}, fog computing \cite{kaur2021systematic}, and efficient edge computing \cite{zhang2021joint} has been continuously researched to optimize the task scheduling system.  \\   \hline

		\small{\textcircled{\small{\footnotesize{7}}}} \footnotesize{Big data underlying basic}   & The underlying technology of big data mainly provides common basic functions such as namespace and configuration service.
		& The framework is mainly Zookeeper, which promotes high potency and high accessibility of data \cite{munjal2022big}. \\ \hline
		
		\small{\textcircled{\small{\footnotesize{8}}}} \footnotesize{Data retrieval}   & With the accumulation of more and more data, except data analysis, some technologies are also in need to achieve fast and complex queries with multiple conditions. Algorithms such as cloud-assisted IoT \cite{wang2021privacy} for data retrieval have also been added gradually to speed up retrieval. & Cloud-assisted IoT data retrieval technology has emerged and can guarantee privacy \cite{wang2021privacy}.\\  \hline
		
		\small{\textcircled{\small{\footnotesize{9}}}} \footnotesize{Big data cluster installation management}       & A complete big data platform should include all kinds of functions mentioned above. If users rely on the operation and maintenance personnel to install each component separately, the workload is relatively large. Cluster installations of big data provide a useful way to quickly install these components \cite{awaysheh2021big}.    & Currently, the most common big data cluster installation management frameworks in the industry include CDH, HDP, CDP, etc. \\  \hline
		\bottomrule
	\end{tabularx}
\end{table*}

\subsection{Metaverse Benefits}

\begin{figure}[h]
	\centering
	\includegraphics[scale = 0.65]{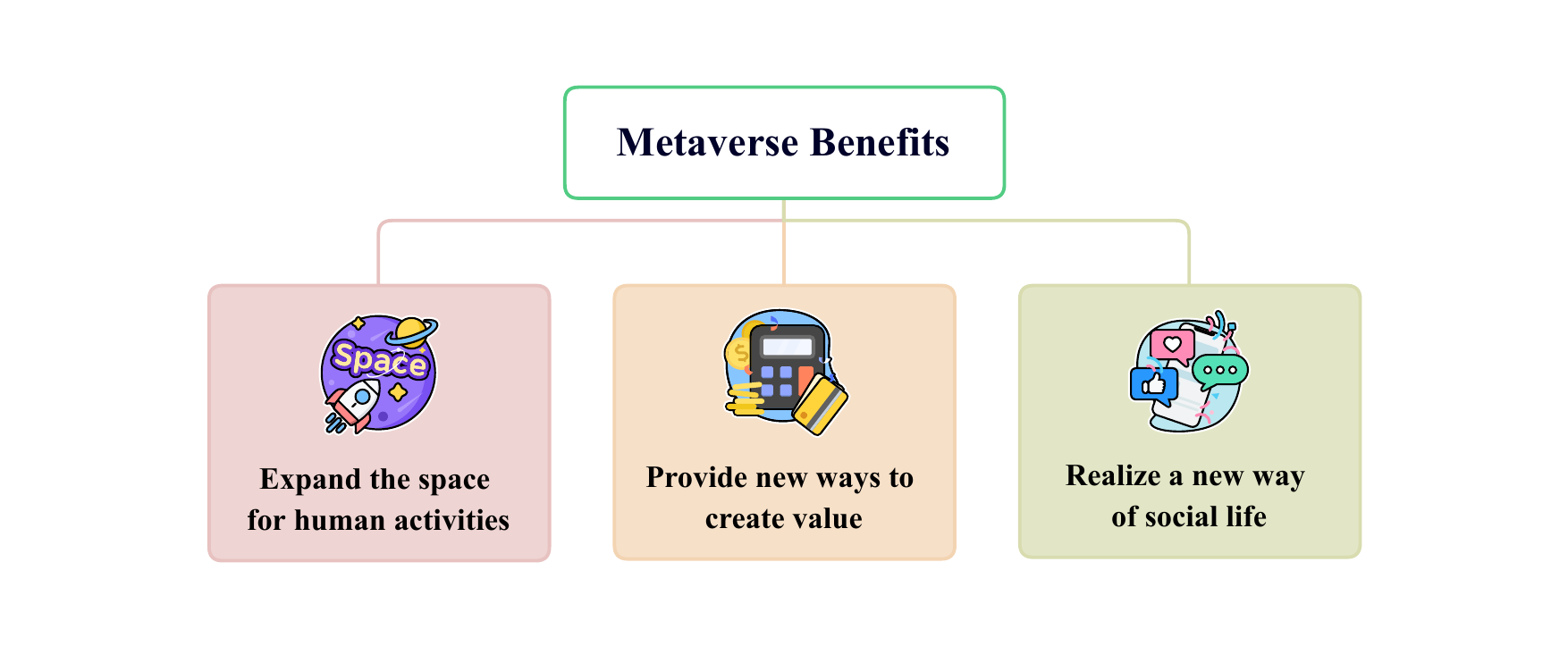}
	\caption{The benefits of the Metaverse}
	\label{fig:Benefit}
\end{figure}

The Metaverse incorporates a variety of emerging technologies with which it innovates, develops, and applies. Integration is the focus of the Metaverse, and its gradual realization and promotion will greatly affect human society. Starting with economic construction, a new financial system and a new business model matching the Metaverse need to be constructed to achieve popularization. The further application of the Metaverse will promote the formulation of new rules and the formation of a new economic system, which will eventually lead society to a new civilization. The benefits of the Metaverse can be summarized as follows, as shown in the Figure \ref{fig:Benefit}.

\textbf{The Metaverse will expand the space for human activities.} The digital twin technology realizes the mapping from the physical world to the digital world, and users can explore and create in the digital world based on the modeling of the real world \cite{tao2018digital,liu2021review}. The development of XR technology offers possibilities from the digital world to the physical world \cite{ratcliffe2021extended}. AR glasses devices such as Microsoft HoloLens, Google Glass \footnote{https://www.microsoft.com/en-us/hololens}, Magic Leap \footnote{https://www.magicleap.com/en-us/}, etc. have begun to be applied in some industries. With the help of these technologies and devices, users switch smoothly between the real world and the virtual world, interacting with elements of both worlds at the same time. In addition, the development of equipment such as drones and intelligent robots has also provided technical support for the strong interaction between the digital world and the physical world. For example, in the event of a disaster, we can quickly and comprehensively analyze the disaster site based on the digital twin in the Metaverse, and directly call various drones and robots to participate in emergency repair and disaster relief activities through the program \cite{fan2021disaster}.

\textbf{The Metaverse will provide new ways to create value.} First, the Metaverse allows for open and free creation. The Metaverse is all-encompassing and cannot be separated from the innovative creations of numerous users. Such a huge content project needs to be dominated by open user creation. User-generated content (UGC) will become digital assets with valuable attributes. Second, the blockchain-based economic system of the Metaverse is safe and stable \cite{lee2021all,yang2022fusing,watson2022virtual}. The economic system of the Metaverse protects users' virtual rights and allows users' virtual assets to flow unimpeded between different platforms. It is interconnected with the real economic system. In the Metaverse, blockchain technology can not only allow digital assets to confirm rights, transfer, and ensure asset security, but also allow digital economic activities in the Metaverse to accumulate and form a large amount of digital wealth. Tokens \cite{lee2019decentralized} are digital assets on the blockchain and will become an asset bridge connecting the physical world and the digital world of the Metaverse.

\textbf{The Metaverse will realize a new way of social life.} Metaverse will provide users with an immersive experience. At present, games, as the most interactive, informative, and immersive content display method, are currently the main content of the Metaverse. These games give players full autonomy to experience digital life, which actually shows the core of digital life, and it is also the prototype of the Metaverse life \cite{lee2021creators,park2022Metaverse}. The application of virtual reality equipment in the Metaverse can revolutionize the previous digital life and enrich the user's experience. Additionally, the Metaverse provides users with unique and verifiable virtual identities. The virtual identity in the Metaverse has the characteristics of consistency and immersion. Metaverse allows users to freely customize the appearance of digital identities and other characteristics to achieve identity uniqueness. The digital identity will be built on the blockchain, and the digital identity and the real identity can be integrated and unified. Blockchain-based digital identities can not only ensure that the identity is fully controlled by the owner, but also ensure identity security, thereby enhancing the credibility of the digital identity \cite{gadekallu2022blockchain,yang2022fusing,wang2022survey}.

\subsection{Metaverse Applications}
\label{subsec:application}

\begin{figure}[h]
	\centering
	\includegraphics[scale = 0.41]{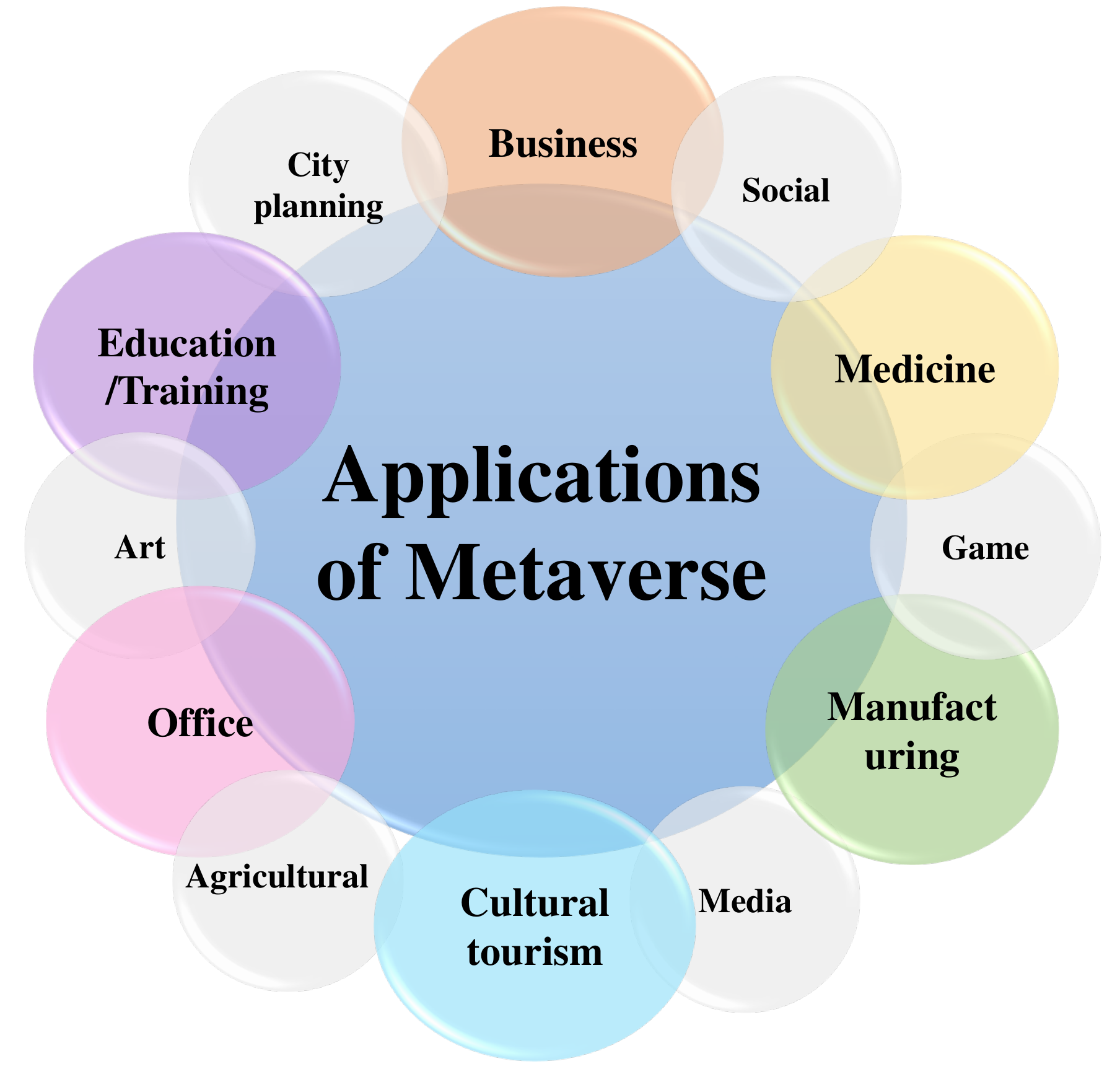}
	\caption{Applications of the Metaverse}
	\label{fig:Application}
\end{figure}

With the continuous development of digital technology, humans may complete the digital migration from the real universe to the Metaverse in the future. With the continuous progress of the Metaverse, the field of application of the Metaverse will become more and more extensive, and it can even be integrated into all aspects of people's lives. Figure \ref{fig:Application} shows some common applications of the Metaverse, and we will discuss five major areas of application in the following sections.

\textbf{Education/Training.} In recent years, the world has been severely affected by the epidemic, and many education methods have shifted from offline to online. The application of Metaverse video in education can immersely display more real scenes and make the learning experience more vivid and interesting. The Metaverse can break the limits of physical distance, allowing teachers or coaches to interact with students in the same virtual space \cite{kye2021educational,mystakidis2022Metaverse,hirsh2022whole}. In the process of teaching, the data of teachers and students will be recorded to improve the quality of class and learning efficiency. Some rigorous training or experimental environments can be easily met in the Metaverse, such as operational instruction for medical students or simulated flight training for pilots \cite{almarzouqi2022prediction}.

\textbf{Business.} The Metaverse expands the path of retail and e-commerce, which meets the convenience of online and offline experiences, providing users with an immersive shopping experience in virtual worlds \cite{jeong2022innovative}. Users can use the digital avatar to try out the product in the virtual world and learn the specific details of the product. In addition, retailers can get timely and useful feedback on product improvements \cite{watson2022virtual,gadalla2013Metaverse}. In addition, marketing and advertising methods have been revolutionized by the Metaverse. Merchants can sell virtual counterparts of real goods and advertise them. Examples of AR experiences on social media clearly identify the potential for future advertising using the Metaverse.

\textbf{Medicine.} Metaverse breaks through the limitations of physical space, and its virtuality, real-time and stability provide conditions for telemedicine. Digital modeling in the virtual world can help doctors understand the patient's condition more comprehensively \cite{werner2022use,mozumder2022overview,chengoden2022Metaverse}. The surgeon will use a robotic arm to perform the operation, enabling an accurate and stable procedure. The Metaverse will also be used to treat mental illness by tricking the brain into a virtual experience, allowing patients to stay happy or erase as many bad memories as possible. In medical education, 3D visualization of organs allows students to have a clearer understanding of the body structure and learn systematic knowledge \cite{skalidis2022cardioverse}. In addition, medical students can operate on virtual patients, thus mastering the skills \cite{locurcio2022dental}.

\textbf{Manufacturing.} The application of the Metaverse in the engineering and design of manufacturing can more truly demonstrate the engineering structure and design principles so that customers can understand the details of the project without leaving home. Engineers and designers can discuss the specific details of products in the virtual space, fully communicate with manufacturers and customers, and timely apply feedback to products, which will improve the efficiency of product design and production \cite{lee2011self,lin2022towards}. For instance, Omniverse\footnote{https://www.nvidia.com/en-us/omniverse/} is an open platform developed by NVIDIA for virtual collaboration and real-time simulation. Designers build content using the platform's 3D tools, developers train AI models in virtual worlds, and engineers build reality-based digital twins.

\textbf{Cultural tourism.} For the cultural tourism industry, Metaverse is a medium that breaks the concept of time and space in tourism and is more like a virtual tourism. The COVID-19 pandemic hinders tourists from visiting tourist attractions in various countries or regions, while this can not happen in Metaverse tourism. Traditional travel is always overcrowded, and time and space prevent people from exploring famous landscapes and monuments around the world. The 3D image of the space is captured digitally and put into the cloud for display. Tourists only need to wear VR glasses to have a ``just go" trip. Various explanations and simulation-style interactions are even more important and help tourists have an immersive experience that goes beyond real-world travel. Matterport\footnote{https://matterport.com/}, a company that makes 3D frames, displayed five Egyptian artifacts in virtual reality. Realistic reproductions based on VR technology are popular with visitors. Visitors can immerse themselves in Egyptian artifacts and learn historical allusions without crossing borders. In addition, the Disney CEO said that the Metaverse is Disney's future. Disney may indeed be the ``happiest place" in the Metaverse when visitors visit the virtual reality version of Disneyland and can talk, dance, and play digitally with various Disney characters.

\textbf{Office.} In the Metaverse, real people are scattered around, but they can gather face-to-face in the virtual scene, which greatly improves the efficiency of work and alleviates the pressure of the urban population \cite{choi2022working}. Some software allows colleagues to meet as avatars in VR or participate in real-world meetings as photo holograms. Currently, collaboration software such as Horizon Workrooms\footnote{https://www.oculus.com/workrooms/} and Microsoft Mesh\footnote{https://www.microsoft.com/en-us/mesh} are available. Remote work has become more common under the influence of the new global epidemic, but traditional remote work still faces some problems, such as a lack of real-time interaction and low communication efficiency. The Metaverse enables virtual offices to be conducted in ``face-to-face" interaction. Google launched the ``Project Starline" program, which aims to enable 3D remote interaction, where participants can observe chat objects from different angles and make physical or eye contact \cite{lawrence2021project}. Facebook has launched Horizon Workrooms, a remote collaboration tool that allows colleagues wearing VR devices to meet face-to-face in the same virtual conference room, breaking the screen.

\section{Making Sense of the Metaverse Through Data Science}
\label{sec:Makesense}

\subsection{The Relationship Between Big Data and Metaverse}

\begin{figure*}[ht]
	\centering
	\includegraphics[scale = 0.34]{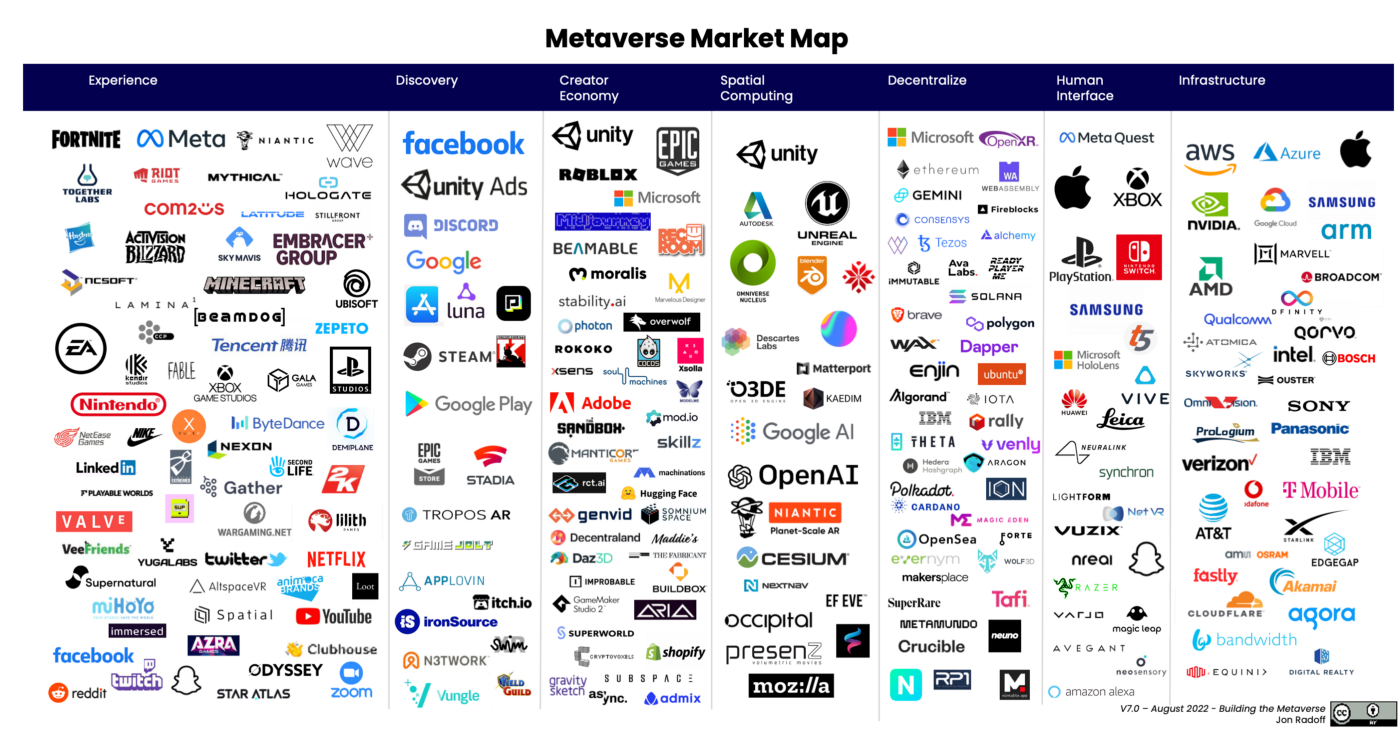}
	\caption{Market map of the Metaverse\protect\footnotemark[8]}
	\label{fig:market}
\end{figure*}

As shown in Fig. \ref{fig:market}, there are many Metaverse applications on the market. From infrastructure and human interface at the bottom, to the middle layer of decentralization, spatial computing, and creator economy, to the application layer of discovery and experience, each layer generates a vast amount of data. Moreover, the communication and interaction between layers depend heavily on data, which will form a huge data network for the flow of information between applications. Therefore, big data technology is the key to the stable operation of the Metaverse market.

\footnotetext[8]{https://medium.com}

\begin{figure}[h]
	\centering
	\includegraphics[scale = 0.38]{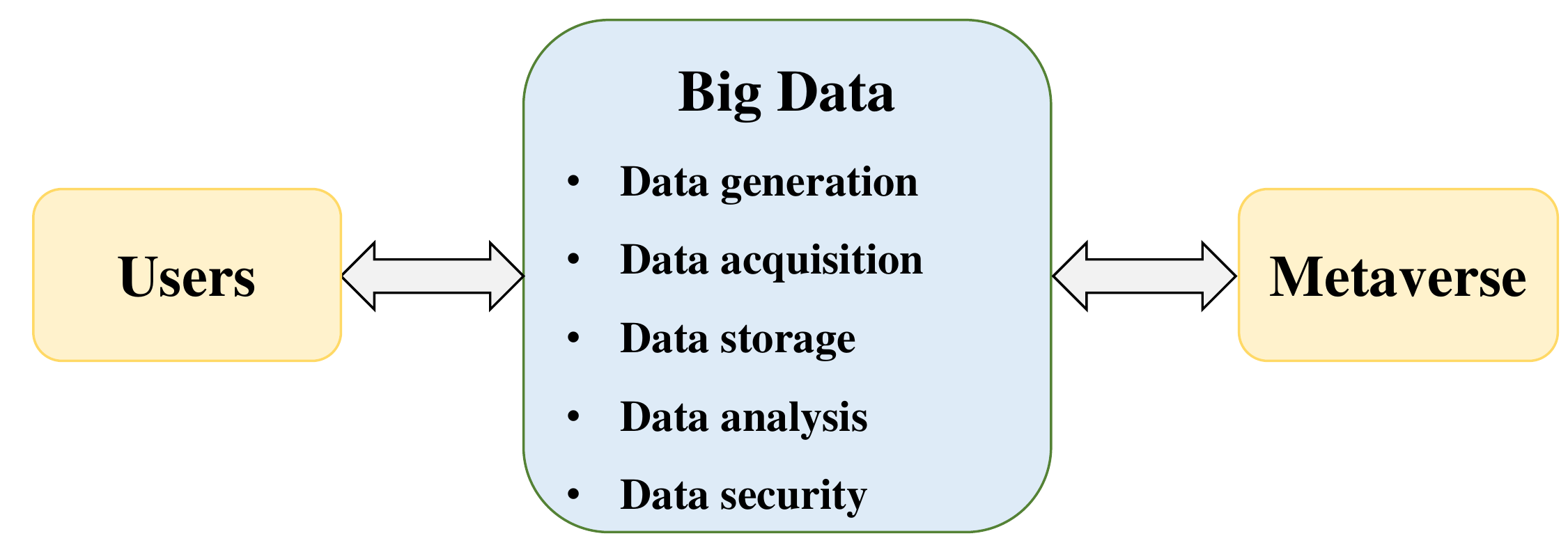}
	\caption{The relationship between big data and the Metaverse}
	\label{fig:Relation}
\end{figure}

The Metaverse is still in its infancy, and there are endless possibilities for what the Metaverse will eventually look like. However, what is certain is that the Metaverse is bound to generate an enormous amount of data. With the possibility of a data explosion, the Metaverse will rely more on data scientists and various big data technologies. Currently, many teams do not have a strategy to maximize their data science capabilities. If these teams are not prepared to deal with rapidly growing data, they may be eliminated by the tide of the Metaverse. Hence, big data and the Metaverse are very closely related. The Metaverse is the spatial structure, scenes, subjects, etc. carved out by digital technology, which is much more digital than the real world. However, in essence, they all exist as data. Big data technology is a bridge between the physical world and the digital world, facilitating the integration of the virtual and the real, as shown in Fig. \ref{fig:Relation}. At present, the digital world and the physical world are merging more and more rapidly, and the Metaverse is also emerging.

Essentially, Web 3.0 describes the potential next stage in the technical aspects of the Internet—a decentralized Internet that runs on top of blockchain technology. And the Metaverse represents a potential next stage at the level of Internet applications. Web 3.0 can actually reshape the value of data. For example, it promises to make data public, transparent and immutable; improve data interoperability; and achieve better value distribution through the token economy. The Metaverse is expected to be a change in the future lifestyle, realizing the symbiosis of virtual and real and more efficient work, social life, and life. Therefore, it is obvious that both Web 3.0 and the Metaverse are closely related to big data. In fact, big data is the medium for the connection between users in the Metaverse, and it is also the basis for users to come into contact with the Metaverse.

At the technical level, the Metaverse can be regarded as a fusion carrier of big data and information technology. Different technologies or hardware are combined, self-looped, and constantly iterated in the Metaverse. 

\subsection{The Big Data Technology Metaverse Needs}

\begin{figure}[h]
	\centering
	\includegraphics[scale = 0.25]{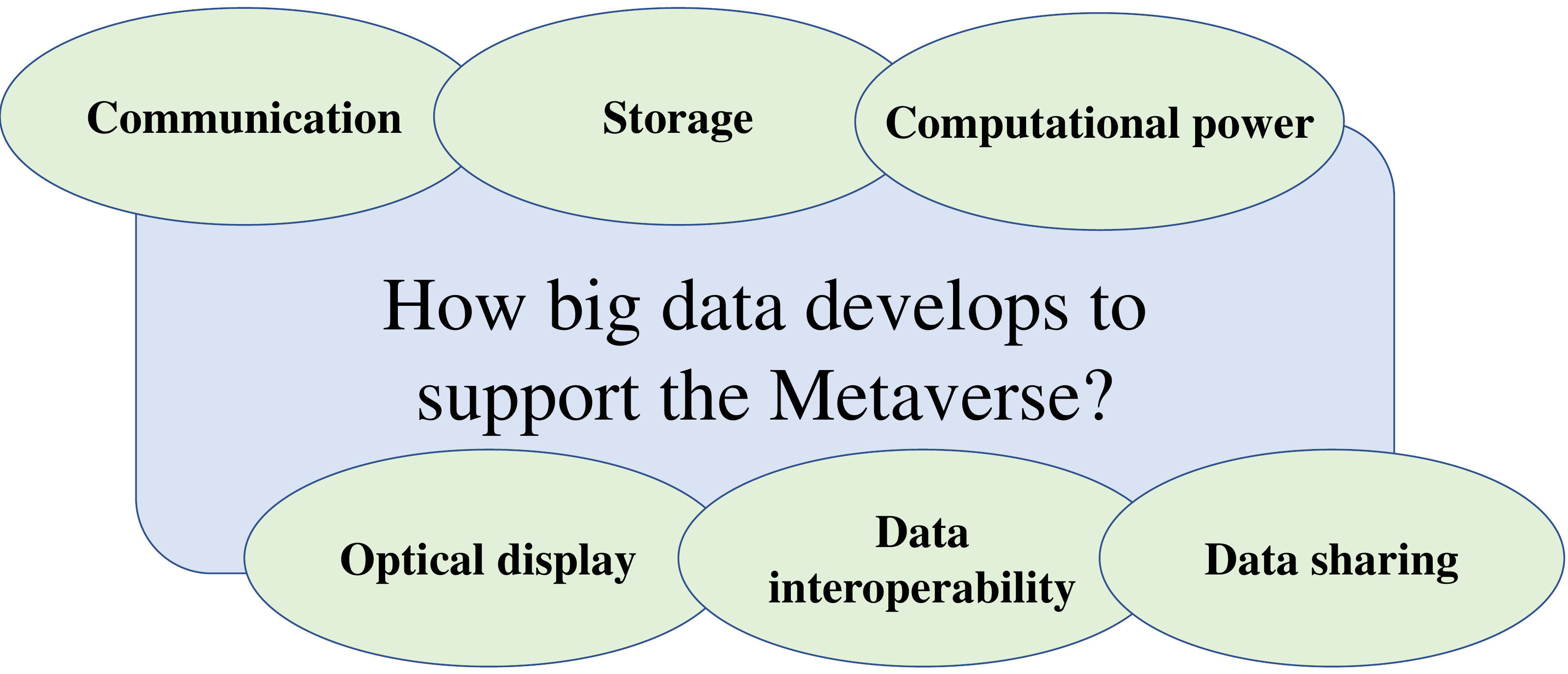}
	\caption{Six aspects of big data develops to support the Metaverse}
	\label{fig:Data_Metaverse}
\end{figure}

According to the above, big data and Metaverse give rise to new technologies or need for some carriers and conditions. We introduce these technologies and conditions below. Big data mainly has the following six technical conditions to face the coming Metaverse, as shown in Fig. \ref{fig:Data_Metaverse}.

\textbf{Communication.} \quad The current technological outcome of combining communication and storage with data is 5G and 6G technology \cite{ning2021survey,gupta2021fusion}. The Metaverse era will inevitably promote all kinds of related software and hardware industries, including big data, cloud computing, blockchain, cybersecurity, latency-sensitive networks, virtual reality, augmented reality, etc. In comparison, 5G has lower latency, faster, and more extensible features than 4G. According to the frequency band, it can be divided into three frequency bands: 24 to 39 GHz, 1 to 6 GHz, and below 1 GHz, which are called high-frequency, mid-frequency, and low-frequency bands, respectively \cite{ni2019research}. High-frequency band 5G can theoretically achieve a maximum throughput of 10 to 20 Gbps, but it only works in a small area. Thus, it is suitable for applications in urban centers \cite{cheng2022will}. Hence, in the 6G era, there will be more opportunities to realize the popularization and application of the Metaverse, because the transmission capacity of 6G may be 100 times higher than that of 5G, up to 1 Tbps. At that time, global coverage and network experience with virtually no lag will be achieved. 6G with AI capabilities is expected to unleash the full potential of radio signals, transforming them into smart radios and providing Metaverse users with a real-time and immersive experience.

\textbf{Storage.} \quad  The traditional centralized storage is to centralize the storage in a complete system. The system includes multiple devices, but there is a unified entrance for data translation. Distributed storage refers to distributing resources on various machines through the network and forming these storage resources into a virtual storage device \cite{chang2008bigtable}, that is, data is stored in a decentralized manner. Compared with centralized storage, distributed storage has the following advantages: easy expansion; high performance; support for hierarchical storage; multi-copy consistency; and storage system standardization \cite{al2010case}. Distributed storage also meets the needs of the decentralization of the Metaverse, and it also encourages people in the Metaverse to be willing to provide protection voluntarily and promotes the realization of a self-running ecology. Furthermore, the explosive growth of digital assets will flood the Metaverse. In addition, applications that require a lot of storage, such as AR (Augmented Reality) and VR (Virtual Reality) streaming, will cause huge storage demands. For example, entry-level VR needs to support 8K resolution and 360° all-round angles, which takes only 20 minutes of storage space. There is no doubt that, in the age of the Metaverse, this need will increase considerably \cite{cai2022compute}. Therefore, the role of big data storage technology in the Metaverse is reflected. Mobile cloud storage and distributed storage systems solve many of the Metaverse data storage problems.

\textbf{Computational power.} \quad After the data input by the Metaverse is stored, it is inevitable to perform computing and analysis \cite{lee2021all}. We have also mentioned above that data continues to grow with the growth of network users and technologies. There is no doubt that the data in the Metaverse is bound to grow massively and rapidly. Traditional computing technologies can no longer meet the current huge demand. Therefore, it makes computing power a very important factor in how data can work in the Metaverse. In other words, the Metaverse has an even more urgent need for computational power \cite{tang2022roadmap}. Hence, the performance improvement of the computing platform and analysis platform for big data provides a guarantee for computational power. With the advent of the 6G era, the service network form with computational power as the core will be ubiquitous on the Internet. At the same time, the synergy between various technologies such as cloud computing, edge computing, and operations in the network will also optimize the utilization of resources. The development of big data computational power can enhance data interaction, which will surely accelerate the formation of the Metaverse \cite{tang2021computing}.

\textbf{Data interoperability.} \quad In the Metaverse, users must create digital avatars to mark their digital footprints. The Metaverse is not a single platform, and users can move their avatars and digital assets across multiple platforms. In order to achieve this seamless user experience, data interoperability must be achieved \cite{gadekallu2022blockchain,lim2022realizing}. That is, the data collected by an entity in the Metaverse should be able to move across platforms and carriers. The Metaverse will contain many sub-Metaverses, equivalent to different digital environments. In the switching and interaction of sub-Metaverses, data needs to have the ability to use and share information across environments, ensuring that consumers have a seamless experience between sub-Metaverses \cite{nawaratne2018self,gonzalez2019interoperability}. The boundaries of various platforms and software have been broken, and users’ digital identities are able to move legally and freely, while digital assets are also more convenient to trade and circulate. It makes the Metaverse an interconnected whole, rather than the currently fragmented internet.

\textbf{Optical display.} \quad It is crucial to the display and presentation of content in the Metaverse. In addition to the underlying support of computational power for the Metaverse mentioned above to ensure the high-speed accuracy of data transmission, ultra HD (high-definition) and AR/VR devices are also important means to help users obtain an excellent immersive experience \cite{bhattacharya2021coalition}. AR/VR, as the interactive medium of the Metaverse, connects the virtual world with the real world \cite{ning2021survey}. AR/VR has higher requirements for image processing and display. Processors, storage, and optical display devices account for a relatively high proportion of the cost, and the industry chain is relatively mature. With the advancement of the Metaverse, the application of voxel modeling will increasingly require hardware such as storage and computing power. With the help of cloud storage and cloud computing, the development of cloud VR and AR can be promoted. This will greatly reduce the requirements for terminal equipment and make the equipment portable and easy to carry. In addition, cloud servers equipped with cloud storage and cloud computing technologies can make data-intensive and computing-intensive tasks more efficient and orderly \cite{tang2021computing}.

\textbf{Data sharing.} \quad Data sharing in the Metaverse can provide useful information to service providers. They use behavior-based data analysis for targeted marketing and advertising delivery to save operational costs \cite{dong2015Secure}. Based on user feedback and product usage data, developers can accurately improve the product. Meanwhile, users can benefit from data sharing. Personalized service and a better user experience are easier to obtain. However, users store a large amount of sensitive and private data in the Metaverse, which easily causes security and privacy issues. Secure data sharing systems need to be proposed for data sharing and information exchange among all parties in the Metaverse. Therefore, the decentralized data management framework based on blockchain is more suitable for Metaverse \cite{chen2019fade,yu2021blockchain}. In addition, data acquired from many IoT devices requires more attention to privacy issues when sharing data \cite{egliston2021critical}. Reasonable data processing and filtering measures should be taken to achieve the purpose of privacy protection.

\section{Security and Privacy of Metaverse Big Data}
\label{sec:Security}

The Metaverse is a 3D virtual space where users can interact with each other in real time. It is made up of individual meta-spaces that can be accessed quickly by any device, anywhere in the world. Since the emergence of the Metaverse, privacy and security issues have been a hot topic \cite{wang2022survey,zhao2022Metaverse,di2021Metaverse}. This problem is pervasive because the Metaverse's cybersecurity infrastructure is in its infancy. While new technologies bring immersive experiences and numerous possibilities, they also bring many security threats. Moreover, due to the mass, concentration, and sensitivity of personal information collected in the Metaverse, once the information is leaked, the loss of personal privacy will be sweeping, which will pose a great threat to the public's personal safety and property safety.

\begin{figure*}[h]
	\centering
	\includegraphics[scale = 0.45]{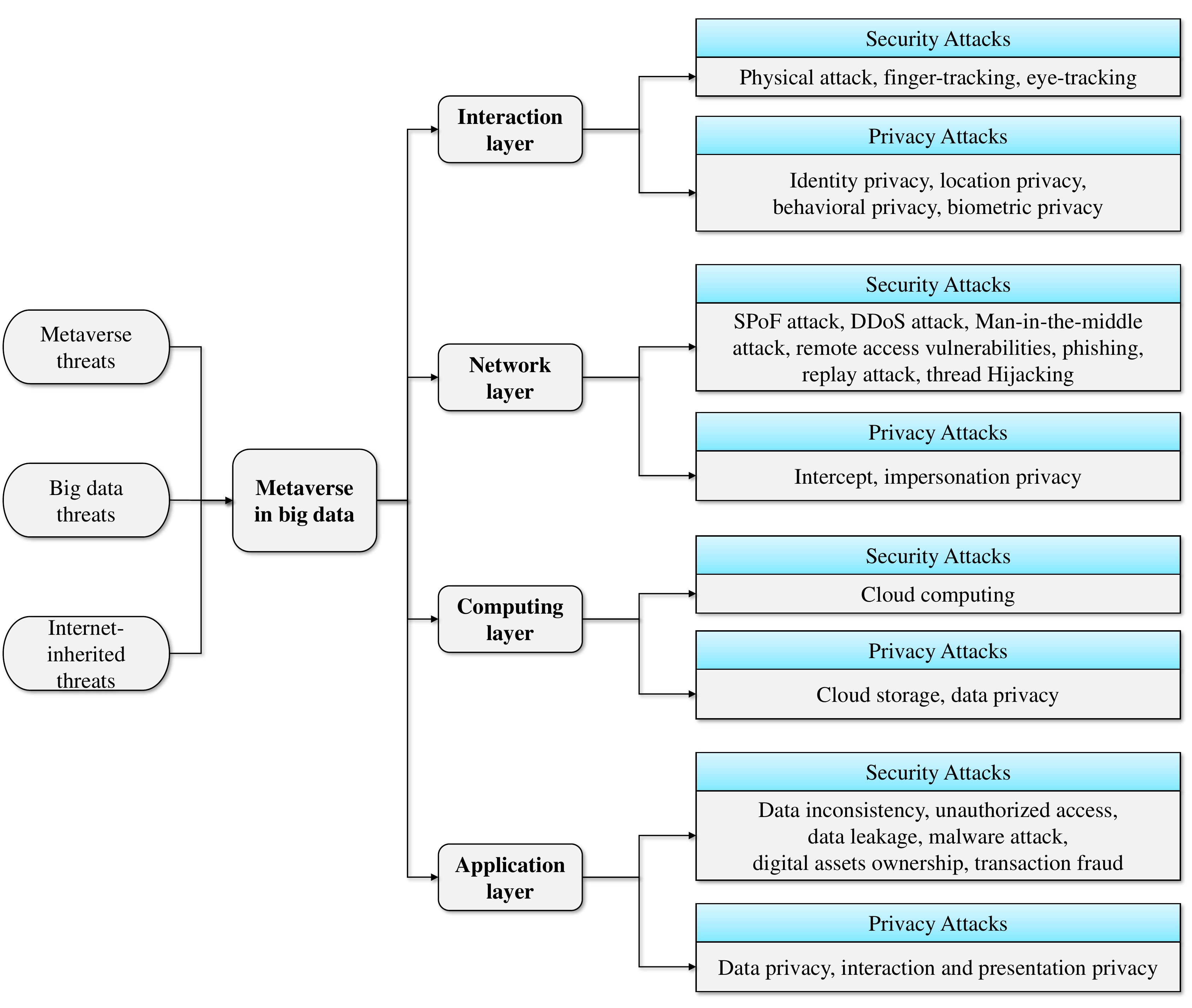}
	\caption{Security and privacy threats and categorization in Metaverse Big Data for different layers}
	\label{fig:SecurityAndPrivacy}
\end{figure*}

Data security and privacy protection are essential in the Metaverse, where everything is connected and easily accessible. Security refers to preventing data from being accessed or destroyed by unauthorized persons. Its main goals are confidentiality, integrity, and availability \cite{von2013information}. Privacy refers to the protection of user identity and sensitive personal information. Security can be defined as protecting data from malicious threats, while privacy is more about the responsible use of data. We will discuss the security and privacy issues and countermeasures in the application of big data technology in the Metaverse from the four levels, as is shown in Fig. \ref{fig:SecurityAndPrivacy}.

\subsection{Security Issues}

The Metaverse generates vast amounts of data, but this data also presents new challenges in security. Security is a very vital issue \cite{cheng2022will}. As the data grows, a series of security issues will inevitably erupt in the Metaverse. The main security issues in the Metaverse include network attacks, technical security flaws, critical infrastructure failures, and so on. If data security is not resolved, the progress of the Metaverse can only be slowed down. It is not advisable to make the Metaverse develop under the premise that data security is not guaranteed. As data assets become larger and larger, the market for buyers will determine the value of the market. In this way, the market for buyers will become larger and larger, and the value of data assets will also increase rapidly. Hence, the security of data will become a critical problem in the coming Metaverse era. Due to the huge number of records, data security must be managed and controlled in layers and at different levels. For example, data can be divided into external public, internal public, secret, confidential, and top secret. Personal data can be divided into ``ordinary" personal data, special personal data, personal data related to business, sensitive data, etc. In the following, we will discuss the security issues in the Metaverse from four layers in Fig. \ref{fig:Framework} that may arise in the Metaverse.

$\bullet $ \textbf{Interaction layer.} Since the Metaverse allows users to self-create and construct parts of this world, much of the data in the interaction layer will come from user-generated content (UGC) \cite{kyto2017augmenting}. It means that the quality and security of data will bring some risks. In addition, mobile threats are more serious than ever. In the interaction layer, the main mobile products are AR and VR devices \cite{carmigniani2011augmented}. With the devices of AR and VR technology, related companies can collect a lot of information about users. This social media is more in-depth. In addition, VR's finger-tracking \cite{shah2014survey} and eye-tracking \cite{clay2019eye} features are also prone to threats to information security such as password theft, information leakage, and so on. If hackers gained access to the information, the security consequences would be disastrous.
	
$\bullet $ \textbf{Network layer.} Since the network is an open public facility, communication is easy to be eavesdropped on, and even a man-in-the-middle attack \cite{conti2016survey,bhushan2017man} can be triggered, so that the communication of devices in the Metaverse is monitored by the attacker, resulting in information leakage. At the same time, DoS attacks \cite{masdari2016survey,zhijun2020low} can be easily carried out in a network environment. Security in the network is quite important because the throughput of data through the network will be greatly increased in the Metaverse era. In the network layer, security is the protection of the underlying network infrastructure from unauthorized access, misuse, or theft \cite{greenberg1998mobile}. With network security, devices, applications, users, and applications are created as a secure infrastructure to work securely. In essence, the protection of network security is also the protection of data in the network, which protects the data in the network from security threats. The main security threats include SPoF, DDoS, MiTM, remote access vulnerabilities, phishing, replay attack, and thread hijacking, etc. \cite{sengupta2020comprehensive}.
	
$\bullet $ \textbf{Computing layer.} In the Metaverse, with the sharp increase in the amount of data, the computing layer requires a very high amount of computing. Therefore, technology is urgently needed to solve this difficult problem. Cloud computing technology is a good choice, and its distributed computing can bring higher computing power \cite{dillon2010cloud, qian2009cloud}. However, it is more difficult to manage securely and more vulnerable to attacks because of its decentralized deployment. The security of cloud computing includes identity and access management, data security, privacy protection, and virtualization security \cite{chen2012data}.
	
$\bullet $ \textbf{Application layer.} The application is a window that opens the data interaction platform to data managers, some digital product manufacturers, and application users. Due to its openness, it is bound to be searched or accessed by unauthorized entities, resulting in data attack behaviors such as data inconsistency, unauthorized access, and data leakage. And most applications for big data clusters are web applications. Therefore, applications in Metaverse Big Data will face the same security risks as those on the normal Internet. When the Metaverse big data application is running on the local AR/VR devices or servers, malware such as viruses and Trojans can invade the system to control or destroy the system, steal or tamper with data, or even launch DoS attacks to paralyze the entire system of Metaverse. At the same time, the Metaverse is also a kind of digital economy, in which big data represents the vast number of digital assets. How to determine and protect the ownership of digital assets and solve the problem of trust in the virtual trading system will affect the development prospects of Metaverse big data. Hence, measures to ensure application security are indispensable.

$\bullet $ \textbf{Scale, real-time and distributed processing}. Due to the huge size of big data, it is relatively difficult to ensure the security of big data in terms of capacity, real-time, distributed architecture and parallel processing. Since there are many technical frameworks involved, and the big data cluster is open, users can communicate with multiple data nodes at the same time, so there is a great potential for security problems.

\subsection{Privacy Issues}

The Metaverse connects many users (devices) together, which will greatly facilitate user interaction. But at the same time, these devices also generate massive amounts of data, posing huge challenges for the development of the Metaverse. Big data is inherently faced with many data privacy issues, and the application of its security model is difficult. In addition, the Metaverse is a complex multi-user interactive system that integrates various technologies, and the protection of data privacy faces more severe challenges. A survey shows that people have a lot of concerns about privacy in the Metaverse. Of those surveyed, 50\% were worried about identity information, 47\% were worried about users being subjected to forced surveillance, and 45\% were concerned about the potential misuse of personal information. With the increasing awareness of security and data privacy issues, many users will mostly consider data privacy issues when using a system or application to decide whether to use the system. If the privacy issue cannot be resolved, the application and popularization of the Metaverse will be greatly affected. Some of the ways the Metaverse may affect user data privacy are as follows:

$\bullet $  \textbf{Interaction layer}. There are many sensing devices in the interaction layer, which are used to collect user behavior data and environmental data. Interactive devices such as XR and brain-computer interfaces (BCI) can also capture sensitive information such as the user's biological data. This extensive data collection poses a threat to users' privacy. The Metaverse needs to provide users with private data management, where users can choose how to authorize and use data. Meanwhile, vulnerable AR/VR devices become gateways for malware intrusions and data breaches. This problem has been extensively seen in VR glasses. The ability to ensure the quality of data generation, promote the virtuous circle of the Metaverse, and protect the data generation process from being illegally obtained by unauthorized parties is crucial to the Metaverse. The main methods to solve this problem are authentication, access control, and falsification of data.
	
$\bullet $ \textbf{Network layer}. When analyzing user behavior, the big data analysis system of the Metaverse usually collects user behavior on the client. The system processes the data (compression and packaging, etc.) and then uploads it to the server for storage and analysis. Data needs to be transmitted through the public network between the client and the server, which brings a series of risks of privacy disclosure. For example, a third party may intercept the transmitted data during the transmission to obtain the user behavior data. This user data reflects some specific user behaviors and contains user privacy. Secure transport protocols, device verification, or encryption of the transmitted content can solve these problems \cite{chen2012data,jian2021hybrid}.
	
$\bullet $ \textbf{Computing layer}. The free flow of data in the ternary world enables digital ecology to finally facilitate the integration of virtual and real worlds \cite{wang2022survey}. The flow of data in the Metaverse is massive and fast. It has all the characteristics of today's Internet. During the transmission process, personal information may be intercepted and identified, posing challenges to data privacy protection. At the same time, data can only have its value after being processed. The big data processing of the Metaverse is not only a challenge, but also one that will promote the development of the Metaverse. On the one hand, data processing is to protect data from being actively leaked, and the other is to obtain meaningful information from it without infringing on user privacy \cite{jain2016big}. The methods of privacy protection during data processing mainly include De-identification \cite{uzuner2007evaluating}, k-anonymity \cite{sweeney2002k,liu2019voting}, L‑diversity \cite{machanavajjhala2007diversity,mehta2019improved}, differential privacy \cite{friedman2010data,guo2021practical}, etc. Breaches of big data storage systems can expose personal information during data processing \cite{jain2016big}. Moreover, in order to prevent SPoF and incentivize creators, Metaverse adopts a decentralized architecture. In a distributed environment, an application may require multiple datasets from different data centers, so it faces the challenge of privacy protection. When the cloud is used for big data storage, the data is no longer completely controlled by the data owner. Outsourcing data storage is risky because cloud servers may not be fully trusted \cite{jain2016big}. Various techniques have been developed to secure cloud storage, such as attribute-based encryption, homomorphic encryption \cite{acar2018survey}, storage path encryption, and integrity verification.
	
$\bullet $  \textbf{Application layer}. Users use avatars to move and socialize in the digital world. Due to the high connectivity between the Metaverse and the real world, avatar data contains a lot of private information about the user's behavior, trajectory, and location. Advertisers will collect user information through avatars, as was evident in Second Life. Once avatar data is misused, users' safety in real life may even be threatened. Users may hesitate to adopt the Metaverse because of privacy concerns. The only way netizens might consider using the Metaverse is by adopting security tools to provide security protection, such as VPNS, antivirus software, phishing protection, etc. Currently, the Metaverse lacks legal documents to protect users' identities. Companies have found tremendous value in collecting, sharing, and using data about their customers or users, especially from social media. Companies in the Metaverse should obtain permission from users to retain their personal data, comply with their privacy policies, and manage the transparency of the data they collect. It is essential to develop legal guidelines on which technical mechanisms and operational mechanisms can be based \cite{tang2022roadmap}. Enterprises should bear legal responsibilities related to the collection, storage, and processing of personal data.

\section{Open Problems and Opportunities} \label{sec:Opportunities}

In the era of big data, especially in the Metaverse, data is ubiquitous, and the proliferation of data will lead to a surge in communication demand, which in turn promotes a surge in demand for language services. Big data technology is a comprehensive technology that reflects the technological nature of society. Metaverse is really a new concept and is related to many domains. In our opinion, the big data-enhancing Metaverse has the following most challenging issues. Here are seven of the most important of the Metaverse with big data.

\subsection{Privacy and Security}

The ever-increasing growth of personal data poses challenges for businesses in the metaverse, an unavoidable but often overlooked issue. With the advent of the metaverse era, the function of linking reality and virtuality in the metaverse will lead to a substantial increase in the online time of users. In the process, more personal data is bound to be generated. At the enterprise or company group level, in order to gain a deeper understanding of users' thinking and behavior, the enterprise or company that is a Metaverse developer will inevitably collect more personal privacy information about users and even continuously monitor users' behavior patterns. There is no doubt that the Metaverse has collected unprecedented amounts of personal data since the development of the Internet. This data directly causes Metaverse to assume particularly high data protection responsibilities and information regulatory risks. In order to prevent the theft of information and misuse of data, strict supervision of data must run through the entire life cycle of data, including storage and management. Moreover, dealing with Metaverse big data requires an efficient and lightweight security and privacy scheme. Traditional security mechanisms, such as RSA, are not suitable for big data security because they can only process small amounts of data. The rationale behind this is that they are not lightweight.

\subsection{Metaverse Big Data Value}

Note that it is the five keys of volume, velocity, variety, veracity, and value that make big data such a huge business. The first things that come to mind when thinking of big data are large organizations with an Internet background, especially technology and social media platforms, such as Google and Facebook. As more and more smart devices with different functions are connected to the Internet, information perception has become ubiquitous, but most of the data generated has low value density and contains a lot of irrelevant information. Therefore, it is necessary to use machine learning, artificial intelligence, and other technologies to conduct predictable and complex analyses of future trends and patterns. Machine learning algorithms can complete the value extraction of data, but the efficiency is not necessarily high. How to construct an efficient value extraction algorithm will be an urgent problem to be solved in Metaverse big data. More importantly, the quality of data determines the value of data, and high-quality data can often generate higher value.

There are several business factors affecting data quality: 1). Business understanding deviation: Inaccurate business descriptions associated with the data, incomplete rules, or inadequate correlation analysis can lead to errors in data modeling. 2). Changes in business processes: From model design and data acquisition to data transfer and data storage, it is closely related to business. Changes in business processes affect every aspect of data processing. If not handled carefully and comprehensively, data quality will be compromised. 3). Irregular data input: There are many sources of data input in the Metaverse, among which the irregular input of users is one of the important factors affecting the quality of data. Common data entry problems, such as case and special characters, are recorded incorrectly. The quality of manually entered data is closely related to the person recording it. 4). There are many business systems: In the past 20 years, many enterprises and government departments have carried out digital transformation and established their own information systems for information management and business management, which leads to the current information island dilemma. These systems are isolated from one another, and the databases are unable to communicate with one another. When technical personnel retrieve data and view business development data, they can only process them one by one, which makes it easy to commit errors and omissions and omit key information, leading to misjudgment of business development. 5). Data falsification: In order to meet the conditions of data evaluation, data managers will modify and process the rich big  data, sometimes affecting the accuracy and authenticity of the rich data.

\subsection{Computing Power}

Computing power is fundamental to the development of big data in the Metaverse. Without the support of the computing power network, the Metaverse is impossible to achieve. What exactly is computing power? In layman's terms, computing power refers to the processing power of data. From mobile phones and PCs to supercomputers, computing power exists in various intelligent hardware devices. Without computing power, there is no normal application of various software and hardware. As the entire society accelerates toward digitization, computing, as a basic digital technology, has become an extension of human capabilities, enabling the digital transformation and upgrading of all walks of life. Computing power is the core component of the digital industry. The Metaverse has entered the 3D Internet era where virtual and reality are integrated, and the demand for computing power has increased exponentially. Relevant predictions show that, according to the concept of the Metaverse, at least 10 times to the 6th power of the current computing power is required.

\subsection{Lightweight Methods}

The Metaverse is accessible anywhere and anytime, and can be accessed through personal computers and mobile devices. However, the highly immersive virtual experience that the Metaverse provides to users is supported by a large amount of human-computer interaction (HCI) \cite{karray2008human}. HCI is a cross-domain research topic, with an emphasis on the design of computer technology. But the most important purpose of its design is about the interaction between people and computers. The lightweight model simplifies the access process and facilitates the embedding of lightweight interactive devices \cite{zhu2022metaaid}. There are some lightweight algorithms for HCI to enhance the experience of users such as a lightweight method for facial expression recognition based on neural network \cite{zhao2020expression}, the lightweight 2d hand pose estimation \cite{santavas2020attention}, and the lightweight fully convolutional neural network for speech emotion recognition \cite{aftab2022light}. In addition, due to the importance of big data security, many new lightweight methods have also been developed to ensure data security, which is critical in the Metaverse. For example, Ma \textit{et al.} \cite{ma2019lightweight} proposed a lightweight approach to privacy and data protection for mobile multimedia security. Deebak \textit{et al.} \cite{deebak2021lightweight} proposed lightweight authentication methods in smart data computing in IoT or Cloud.

\subsection{Human-Centered Metaverse}

Metaverse big data can transform business intelligence. As the Metaverse grows, businesses or individuals will be able to use cloud data to collect and analyze vast amounts of data from within the platform and from third-party sources to gain rich, actionable insights into audiences and their collective interests and intentions. As people move away from avatar keyboards in immersive virtual environments, we are likely to see a dramatic increase in the reliance on big data analytics in building predictive models and decision-making activities. This leads to a human-centered Metaverse. The Metaverse is the ``universe" of people, and everyone will become a data center node. Baszucki believes that the Metaverse needs to have at least eight elements: identity, friends, immersion, low latency, diversity, anywhere, economic system, and civilization. Basically, these elements are discussed among people. The Metaverse can be understood as a world parallel to the real world and created by people in the real world. For achieving human-centered Metaverse, how to make Metaverse for social good is a research field worthy of in-depth exploration, sustainable and beneficial to mankind.

\subsection{Applications}

In our opinion, there are many open problems and opportunities when developing Metaverse with big data in the future. The best form of data application is the Metaverse itself, and at present, it may be a digital society in which digital life self-reproduces and grows. To address the aforementioned applications in Section \ref{subsec:application}, many technologies and solutions have come into existence. Let us discuss smart cities, which is one of the key applications. At the technical level, smart cities include IoT technology, 5G, video analysis, AI, blockchain, cloud services, 3D visualization technology, etc., all of which are technical applications. The core idea is to empower urban governance through various technologies. The so-called digital twins, CIM \cite{stojanovski2020city,xue2021semantic}, BIM \cite{fu2020bim}, etc., can be simply regarded as part of the Metaverse. In the digital twin city, through 3D technology, a complete urban information model is established, such as population, buildings, underground space, road traffic, infrastructure, lifelines, and other facilities, and the status of all elements can be obtained and detected in real time through IoT perception. It is mapped to the urban information model, and elements such as population, economy, and events are superimposed on the information model through spatio-temporal big data processing. These form a virtual world that mirrors the real world in real time and constitute the content of the virtual world. But at present, twin cities are still dominated by information presentation, and there is still a long way to go in terms of immersion, social interaction, low latency, and economic systems.

\subsection{Web 3.0}

With the gradual increase of Metaverse participants and the precipitation and accumulation of data, the value of data will increase significantly, but how to use data to generate greater value while better ensuring privacy protection is an important issue. Web 3.0 \cite{barassi2012does,rudman2016defining,kshetri2022web} is a decentralized online ecosystem built on the blockchain, which has three features: decentralization, permission-free, and security. Web 3.0 breaks the data island, returns data rights to individual users, and can combine and interact with applications at will. The data of Web 3.0 is stored in the blockchain, and no single system can access all the data. Moreover, Web 3.0 can provide users with more convenient services under the condition of satisfying privacy protection. Users can access certain services without disclosing their personal information. Due to its decentralized nature, it is more difficult for hackers to attack specific databases, and the security is greatly improved compared to Web 2.0 \cite{murugesan2007understanding,o2007web}. In addition, building a decentralized reputation system through multidimensional data vectors makes it possible for various credit-based systems, such as virtual finance.

\section{Conclusions} \label{sec:conclusion}

As a revival topic in 2021, the Metaverse is under revolution in the long term. Both big data and the Metaverse are a revolution for our society. How much big data enhances the Metaverse? It's crucial to have the right Metaverse technology to handle big data. However, they also introduce a significant threat to our privacy. In this survey, we provide a comprehensive review of how Metaverse is changing big data and how the future of Metaverse meets big data. First, we briefly introduce the key concepts of big data and Metaverse, as well as their benefits. Second, we review how to make sense of the Metaverse technology through big data in detail, including the relationship between big data and the Metaverse, the key tasks, and technical requirements when the Metaverse meets big data. Third, we focus on the security and privacy issues, as well as existing countermeasures, that arise when big data collides with the Metaverse, providing a comprehensive overview for related researchers. Finally, some important open problems and opportunities are discussed in detail. We hope this in-depth systematic survey will give a detailed explanation of the relationship between Metaverse and big data and also provide some useful research directions for future study.


\ifCLASSOPTIONcompsoc
\else
\fi

\section*{Acknowledgment}

This research was supported in part by the National Natural Science Foundation of China (Grant Nos. 62002136 and 62272196), Natural Science Foundation of Guangdong Province (Grant No. 2022A1515011861), Guangzhou Basic and Applied Basic Research Foundation (Grant No. 202102020277), and the Young Scholar Program of Pazhou Lab (Grant No. PZL2021KF0023).

\ifCLASSOPTIONcaptionsoff
\newpage
\fi

\bibliographystyle{IEEEtran}
\bibliography{paper.bib}

\begin{thebibliography}{100}
\providecommand{\url}[1]{#1}
\csname url@samestyle\endcsname
\providecommand{\newblock}{\relax}
\providecommand{\bibinfo}[2]{#2}
\providecommand{\BIBentrySTDinterwordspacing}{\spaceskip=0pt\relax}
\providecommand{\BIBentryALTinterwordstretchfactor}{4}
\providecommand{\BIBentryALTinterwordspacing}{\spaceskip=\fontdimen2\font plus
\BIBentryALTinterwordstretchfactor\fontdimen3\font minus
  \fontdimen4\font\relax}
\providecommand{\BIBforeignlanguage}[2]{{%
\expandafter\ifx\csname l@#1\endcsname\relax
\typeout{** WARNING: IEEEtran.bst: No hyphenation pattern has been}%
\typeout{** loaded for the language `#1'. Using the pattern for}%
\typeout{** the default language instead.}%
\else
\language=\csname l@#1\endcsname
\fi
#2}}
\providecommand{\BIBdecl}{\relax}
\BIBdecl

\bibitem{stephenson2003snow}
N.~Stephenson, \emph{Snow crash: A novel}.\hskip 1em plus 0.5em minus
  0.4em\relax Spectra, 2003.

\bibitem{zhang2021study}
C.~Zhang and Y.~Lu, ``Study on artificial intelligence: The state of the art
  and future prospects,'' \emph{Journal of Industrial Information Integration},
  vol.~23, p. 100224, 2021.

\bibitem{huynh2022artificial}
T.~Huynh~The, Q.~V. Pham, X.~Q. Pham, T.~T. Nguyen, Z.~Han, and D.~S. Kim,
  ``Artificial intelligence for the metaverse: A survey,'' \emph{arXiv preprint
  arXiv:2202.10336}, 2022.

\bibitem{ratcliffe2021extended}
J.~Ratcliffe, F.~Soave, N.~Bryan-Kinns, L.~Tokarchuk, and I.~Farkhatdinov,
  ``Extended reality {(XR)} remote research: a survey of drawbacks and
  opportunities,'' in \emph{Proceedings of the CHI Conference on Human Factors
  in Computing Systems}, 2021, pp. 1--13.

\bibitem{xi2022challenges}
N.~Xi, J.~Chen, F.~Gama, M.~Riar, and J.~Hamari, ``The challenges of entering
  the metaverse: An experiment on the effect of extended reality on workload,''
  \emph{Information Systems Frontiers}, pp. 1--22, 2022.

\bibitem{mohanta2019blockchain}
B.~K. Mohanta, D.~Jena, S.~S. Panda, and S.~Sobhanayak, ``Blockchain
  technology: A survey on applications and security privacy challenges,''
  \emph{Internet of Things}, vol.~8, p. 100107, 2019.

\bibitem{gadekallu2022blockchain}
T.~R. Gadekallu, T.~Huynh~The, W.~Wang, G.~Yenduri, P.~Ranaweera, Q.~V. Pham,
  D.~B. da~Costa, and M.~Liyanage, ``Blockchain for the metaverse: A review,''
  \emph{arXiv preprint arXiv:2203.09738}, 2022.

\bibitem{lee2021all}
L.-H. Lee, T.~Braud, P.~Zhou, L.~Wang, D.~Xu, Z.~Lin, A.~Kumar, C.~Bermejo, and
  P.~Hui, ``All one needs to know about metaverse: A complete survey on
  technological singularity, virtual ecosystem, and research agenda,''
  \emph{arXiv preprint arXiv:2110.05352}, 2021.

\bibitem{du2018big}
M.~Du, K.~Wang, Y.~Chen, X.~Wang, and Y.~Sun, ``Big data privacy preserving in
  multi-access edge computing for heterogeneous internet of things,''
  \emph{IEEE Communications Magazine}, vol.~56, no.~8, pp. 62--67, 2018.

\bibitem{ooi2022sense}
B.~C. Ooi, K.-L. Tan, A.~Tung, G.~Chen, M.~Z. Shou, X.~Xiao, and M.~Zhang,
  ``Sense the physical, walkthrough the virtual, manage the metaverse: A
  data-centric perspective,'' \emph{arXiv preprint arXiv:2206.10326}, 2022.

\bibitem{mohammadi2018deep}
M.~Mohammadi, A.~Al-Fuqaha, S.~Sorour, and M.~Guizani, ``Deep learning for
  {IoT} big data and streaming analytics: A survey,'' \emph{IEEE Communications
  Surveys \& Tutorials}, vol.~20, no.~4, pp. 2923--2960, 2018.

\bibitem{han2022dynamic}
Y.~Han, D.~Niyato, C.~Leung, C.~Miao, and D.~I. Kim, ``A dynamic resource
  allocation framework for synchronizing metaverse with iot service and data,''
  in \emph{IEEE International Conference on Communications}.\hskip 1em plus
  0.5em minus 0.4em\relax IEEE, 2022, pp. 1196--1201.

\bibitem{cai2022compute}
Y.~Cai, J.~Llorca, A.~M. Tulino, and A.~F. Molisch, ``Compute-and
  data-intensive networks: The key to the {Metaverse},'' \emph{arXiv preprint
  arXiv:2204.02001}, 2022.

\bibitem{sagiroglu2013Big}
S.~Sagiroglu and D.~Sinanc, ``Big data: A review,'' in \emph{International
  Conference on Collaboration Technologies and Systems}.\hskip 1em plus 0.5em
  minus 0.4em\relax IEEE, 2013, pp. 42--47.

\bibitem{hariri2019uncertainty}
R.~H. Hariri, E.~M. Fredericks, and K.~M. Bowers, ``Uncertainty in big data
  analytics: survey, opportunities, and challenges,'' \emph{Journal of Big
  Data}, vol.~6, no.~1, pp. 1--16, 2019.

\bibitem{ge2018big}
M.~Ge, H.~Bangui, and B.~Buhnova, ``Big data for internet of things: a
  survey,'' \emph{Future Generation Computer Systems}, vol.~87, pp. 601--614,
  2018.

\bibitem{katal2013big}
A.~Katal, M.~Wazid, and R.~H. Goudar, ``Big data: issues, challenges, tools and
  good practices,'' in \emph{Sixth International Conference on Contemporary
  Computing}.\hskip 1em plus 0.5em minus 0.4em\relax IEEE, 2013, pp. 404--409.

\bibitem{ning2021survey}
H.~Ning, H.~Wang, Y.~Lin, W.~Wang, S.~Dhelim, F.~Farha, J.~Ding, and
  M.~Daneshmand, ``A survey on metaverse: the state-of-the-art, technologies,
  applications, and challenges,'' \emph{arXiv preprint arXiv:2111.09673}, 2021.

\bibitem{hajjaji2021big}
Y.~Hajjaji, W.~Boulila, I.~R. Farah, I.~Romdhani, and A.~Hussain, ``Big data
  and {IoT}-based applications in smart environments: A systematic review,''
  \emph{Computer Science Review}, vol.~39, p. 100318, 2021.

\bibitem{sun2022matrix}
X.~Sun, Y.~Lu, J.~Sun, B.~Tang, K.~D. Rehak, and S.~Zhang, ``Matrix syncer--a
  multi-chain data aggregator for supporting blockchain-based metaverses,''
  \emph{arXiv preprint arXiv:2204.04272}, 2022.

\bibitem{park2022method}
D.~Park, J.~M. Kim, J.~Jung, and S.~Choi, ``Method to create a metaverse using
  smartphone data,'' in \emph{International Conference on Human-Computer
  Interaction}.\hskip 1em plus 0.5em minus 0.4em\relax Springer, 2022, pp.
  45--57.

\bibitem{yang2022smart}
Y.~Yang, K.~Siau, W.~Xie, and Y.~Sun, ``Smart health intelligent healthcare
  systems in the metaverse, artificial intelligence, and data science era,''
  \emph{Journal of Organizational and End User Computing}, vol.~34, no.~1, pp.
  1--14, 2022.

\bibitem{angelini2022towards}
L.~Angelini, M.~Mecella, H.-N. Liang, M.~Caon, E.~Mugellini, O.~Abou~Khaled,
  and D.~Bernardini, ``Towards an emotionally augmented metaverse: a framework
  for recording and analysing physiological data and user behaviour,'' in
  \emph{13th Augmented Human International Conference}, 2022, pp. 1--5.

\bibitem{el2018investigating}
H.~El~Bousty, S.-d. Krit, M.~Elasikri, H.~Dani, K.~Karimi, K.~Bendaoud, and
  M.~Kabrane, ``Investigating business intelligence in the era of big data:
  Concepts, benefits and challenges,'' in \emph{Proceedings of the Fourth
  International Conference on Engineering \& MIS}, 2018, pp. 1--9.

\bibitem{bryson1999visually}
S.~Bryson, D.~Kenwright, M.~Cox, D.~Ellsworth, and R.~Haimes, ``Visually
  exploring gigabyte data sets in real time,'' \emph{Communications of the
  ACM}, vol.~42, no.~8, pp. 82--90, 1999.

\bibitem{russom2011big}
P.~Russom \emph{et~al.}, ``Big data analytics,'' \emph{TDWI Best Practices
  Report, Fourth Quarter}, vol.~19, no.~4, pp. 1--34, 2011.

\bibitem{park2022Metaverse}
S.-M. Park and Y.-G. Kim, ``A metaverse: Taxonomy, components, applications,
  and open challenges,'' \emph{IEEE Access}, vol.~10, pp. 4209--4251, 2022.

\bibitem{wang2022survey}
Y.~Wang, Z.~Su, N.~Zhang, D.~Liu, R.~Xing, T.~H. Luan, and X.~Shen, ``A survey
  on metaverse: Fundamentals, security, and privacy,'' \emph{arXiv preprint
  arXiv:2203.02662}, 2022.

\bibitem{di2021Metaverse}
R.~Di~Pietro and S.~Cresci, ``Metaverse: Security and privacy issues,'' in
  \emph{Third IEEE International Conference on Trust, Privacy and Security in
  Intelligent Systems and Applications}.\hskip 1em plus 0.5em minus 0.4em\relax
  IEEE, 2021, pp. 281--288.

\bibitem{zhao2022Metaverse}
R.~Zhao, Y.~Zhang, Y.~Zhu, R.~Lan, and Z.~Hua, ``Metaverse: Security and
  privacy concerns,'' \emph{arXiv preprint arXiv:2203.03854}, 2022.

\bibitem{yang2022fusing}
Q.~Yang, Y.~Zhao, H.~Huang, Z.~Xiong, J.~Kang, and Z.~Zheng, ``Fusing
  blockchain and ai with metaverse: A survey,'' \emph{IEEE Open Journal of the
  Computer Society}, 2022.

\bibitem{jeon2022blockchain}
H.~J. Jeon, H.~C. Youn, S.~M. Ko, and T.~H. Kim, ``Blockchain and ai meet in
  the metaverse,'' \emph{Advances in the Convergence of Blockchain and
  Artificial Intelligence}, p.~73, 2022.

\bibitem{al2019big}
Z.~A. Al-Sai, R.~Abdullah \emph{et~al.}, ``Big data impacts and challenges: a
  review,'' in \emph{IEEE Jordan International Joint Conference on Electrical
  Engineering and Information Technology}.\hskip 1em plus 0.5em minus
  0.4em\relax IEEE, 2019, pp. 150--155.

\bibitem{younas2019research}
M.~Younas, ``Research challenges of big data,'' \emph{Service Oriented
  Computing and Applications}, vol.~13, no.~2, pp. 105--107, 2019.

\bibitem{mhammedi2021heterogeneous}
S.~Mhammedi and N.~Gherabi, ``Heterogeneous integration of big data using
  semantic {Web} technologies,'' in \emph{Intelligent Systems in Big Data,
  Semantic Web and Machine Learning}.\hskip 1em plus 0.5em minus 0.4em\relax
  Springer, 2021, pp. 167--177.

\bibitem{rusu2013converting}
O.~Rusu, I.~Halcu, O.~Grigoriu, G.~Neculoiu, V.~Sandulescu, M.~Marinescu, and
  V.~Marinescu, ``Converting unstructured and semi-structured data into
  knowledge,'' in \emph{11th RoEduNet International Conference}.\hskip 1em plus
  0.5em minus 0.4em\relax IEEE, 2013, pp. 1--4.

\bibitem{kumar2021integrated}
K.~Kumar, ``Integrated benchmarking standard and decision support system for
  structured, semi structured, unstructured retail data,'' \emph{Wireless
  Networks}, pp. 1--11, 2021.

\bibitem{agrawal2011challenges}
D.~Agrawal, P.~Bernstein, E.~Bertino, S.~Davidson, U.~Dayal, M.~Franklin,
  J.~Gehrke, L.~Haas, A.~Halevy, J.~Han \emph{et~al.}, ``Challenges and
  opportunities with big data 2011-1,'' 2011.

\bibitem{oguntimilehin2014review}
A.~Oguntimilehin and E.-O. Ademola, ``A review of big data management, benefits
  and challenges,'' \emph{A Review of Big Data Management, Benefits and
  Challenges}, vol.~5, no.~6, pp. 1--7, 2014.

\bibitem{choi2018big}
T.-M. Choi, S.~W. Wallace, and Y.~Wang, ``Big data analytics in operations
  management,'' \emph{Production and Operations Management}, vol.~27, no.~10,
  pp. 1868--1883, 2018.

\bibitem{grover2018creating}
V.~Grover, R.~H. Chiang, T.~P. Liang, and D.~Zhang, ``Creating strategic
  business value from big data analytics: A research framework,'' \emph{Journal
  of Management Information Systems}, vol.~35, no.~2, pp. 388--423, 2018.

\bibitem{sarker2021data}
I.~H. Sarker, ``Data science and analytics: an overview from data-driven smart
  computing, decision-making and applications perspective,'' \emph{SN Computer
  Science}, vol.~2, no.~5, pp. 1--22, 2021.

\bibitem{zhong2016visualization}
R.~Y. Zhong, S.~Lan, C.~Xu, Q.~Dai, and G.~Q. Huang, ``Visualization of
  rfid-enabled shopfloor logistics big data in cloud manufacturing,'' \emph{The
  International Journal of Advanced Manufacturing Technology}, vol.~84, no.~1,
  pp. 5--16, 2016.

\bibitem{nobre2017scientific}
G.~C. Nobre and E.~Tavares, ``Scientific literature analysis on big data and
  internet of things applications on circular economy: a bibliometric study,''
  \emph{Scientometrics}, vol. 111, no.~1, pp. 463--492, 2017.

\bibitem{aceto2020industry}
G.~Aceto, V.~Persico, and A.~Pescap{\'e}, ``Industry 4.0 and health: Internet
  of things, big data, and cloud computing for healthcare 4.0,'' \emph{Journal
  of Industrial Information Integration}, vol.~18, p. 100129, 2020.

\bibitem{zheng2020commerce}
K.~Zheng, Z.~Zhang, and B.~Song, ``E-commerce logistics distribution mode in
  big-data context: a case analysis of jd. com,'' \emph{Industrial Marketing
  Management}, vol.~86, pp. 154--162, 2020.

\bibitem{jiang2018rethinking}
D.~Jiang, L.~Huo, and H.~Song, ``Rethinking behaviors and activities of base
  stations in mobile cellular networks based on big data analysis,'' \emph{IEEE
  Transactions on Network Science and Engineering}, vol.~7, no.~1, pp. 80--90,
  2018.

\bibitem{yanhua2020application}
Z.~Yanhua, ``The application of artificial intelligence in foreign language
  teaching,'' in \emph{International Conference on Artificial Intelligence and
  Education}.\hskip 1em plus 0.5em minus 0.4em\relax IEEE, 2020, pp. 40--42.

\bibitem{cao2021innovation}
Y.~Cao, ``Innovation and reform of accounting professional training model based
  on the artificial intelligence,'' in \emph{Journal of Physics: Conference
  Series}, vol. 1915, no.~4.\hskip 1em plus 0.5em minus 0.4em\relax IOP
  Publishing, 2021, p. 042023.

\bibitem{kim2021advertising}
J.~Kim, ``Advertising in the metaverse: Research agenda,'' \emph{Journal of
  Interactive Advertising}, vol.~21, no.~3, pp. 141--144, 2021.

\bibitem{parthasarathy2022web}
A.~Parthasarathy, ``{Web} 3.0 \& metaverse: Is this the future of the
  internet?'' 2022.

\bibitem{cook2020spatial}
A.~V. Cook, M.~Bechtel, S.~Anderson, D.~R. Novak, N.~Nodi, and J.~Parekh, ``The
  spatial {Web} and {Web} 3.0: What business leaders should know about the next
  era of computing,'' \emph{Deloitte Insights}, 2020.

\bibitem{singh2015survey}
D.~Singh and C.~K. Reddy, ``A survey on platforms for big data analytics,''
  \emph{Journal of Big Data}, vol.~2, no.~1, pp. 1--20, 2015.

\bibitem{wang2022metasocieties}
F.-Y. Wang, R.~Qin, X.~Wang, and B.~Hu, ``Metasocieties in metaverse:
  Metaeconomics and metamanagement for metaenterprises and metacities,''
  \emph{IEEE Transactions on Computational Social Systems}, vol.~9, no.~1, pp.
  2--7, 2022.

\bibitem{qi2020overview}
J.~Qi, P.~Yang, L.~Newcombe, X.~Peng, Y.~Yang, and Z.~Zhao, ``An overview of
  data fusion techniques for internet of things enabled physical activity
  recognition and measure,'' \emph{Information Fusion}, vol.~55, pp. 269--280,
  2020.

\bibitem{salo2018users}
M.~Salo and M.~Makkonen, ``Why do users switch mobile applications?: Trialing
  behavior as a predecessor of switching behavior,'' \emph{Communications of
  the Association for Information Systems}, vol.~42, 2018.

\bibitem{luceri2018vivo}
L.~Luceri, F.~Cardoso, M.~Papandrea, S.~Giordano, J.~Buwaya, S.~Kundig, C.~M.
  Angelopoulos, J.~Rolim, Z.~Zhao, J.~L. Carrera \emph{et~al.}, ``Vivo: A
  secure, privacy-preserving, and real-time crowd-sensing framework for the
  internet of things,'' \emph{Pervasive and Mobile Computing}, vol.~49, pp.
  126--138, 2018.

\bibitem{ribeiro2021towards}
S.~Ribeiro-Navarrete, J.~R. Saura, and D.~Palacios-Marqu{\'e}s, ``Towards a new
  era of mass data collection: Assessing pandemic surveillance technologies to
  preserve user privacy,'' \emph{Technological Forecasting and Social Change},
  vol. 167, p. 120681, 2021.

\bibitem{feng2022secure}
C.~Feng, G.~P. Duan, D.~You, and X.~W. Zhang, ``Secure data sharing solution
  for mobile cloud storage,'' in \emph{International Conference on Cyber
  Security, Artificial Intelligence, and Digital Economy}, vol. 12330.\hskip
  1em plus 0.5em minus 0.4em\relax SPIE, 2022, pp. 60--63.

\bibitem{mijuskovic2021resource}
A.~Mijuskovic, A.~Chiumento, R.~Bemthuis, A.~Aldea, and P.~Havinga, ``Resource
  management techniques for cloud/fog and edge computing: An evaluation
  framework and classification,'' \emph{Sensors}, vol.~21, no.~5, p. 1832,
  2021.

\bibitem{wadhwa2022optimized}
H.~Wadhwa and R.~Aron, ``Optimized task scheduling and preemption for
  distributed resource management in fog-assisted {IoT} environment,''
  \emph{The Journal of Supercomputing}, pp. 1--39, 2022.

\bibitem{chintapalli2016benchmarking}
S.~Chintapalli, D.~Dagit, B.~Evans, R.~Farivar, T.~Graves, M.~Holderbaugh,
  Z.~Liu, K.~Nusbaum, K.~Patil, B.~J. Peng \emph{et~al.}, ``Benchmarking
  streaming computation engines: Storm, flink and spark streaming,'' in
  \emph{IEEE International Parallel and Distributed Processing Symposium
  Workshops}.\hskip 1em plus 0.5em minus 0.4em\relax IEEE, 2016, pp.
  1789--1792.

\bibitem{kumar2021evolution}
M.~Kumar, K.~Dubey, and R.~Pandey, ``Evolution of emerging computing paradigm
  cloud to fog: applications, limitations and research challenges,'' in
  \emph{11th International Conference on Cloud Computing, Data Science \&
  Engineering}.\hskip 1em plus 0.5em minus 0.4em\relax IEEE, 2021, pp.
  257--261.

\bibitem{xie2022research}
C.~Xie, Q.~Hua, J.~Zhao, R.~Guo, H.~Yao, and L.~Guo, ``Research on energy
  saving technology at mobile edge networks of {IoTs} based on big data
  analysis,'' \emph{Complex \& Intelligent Systems}, pp. 1--10, 2022.

\bibitem{li2022big}
X.~Li, H.~Liu, W.~Wang, Y.~Zheng, H.~Lv, and Z.~Lv, ``Big data analysis of the
  internet of things in the digital twins of smart city based on deep
  learning,'' \emph{Future Generation Computer Systems}, vol. 128, pp.
  167--177, 2022.

\bibitem{balcerzak2022blockchain}
A.~P. Balcerzak, E.~Nica, E.~Rogalska, M.~Poliak, T.~Klie{\v{s}}tik, and O.-M.
  Sabie, ``Blockchain technology and smart contracts in decentralized
  governance systems,'' \emph{Administrative Sciences}, vol.~12, no.~3, p.~96,
  2022.

\bibitem{chen2022influence}
M.~Chen, ``The influence of big data analysis of intelligent manufacturing
  under machine learning on start-ups enterprise,'' \emph{Enterprise
  Information Systems}, vol.~16, no.~2, pp. 347--362, 2022.

\bibitem{ibrahim2021task}
I.~M. Ibrahim \emph{et~al.}, ``Task scheduling algorithms in cloud computing: A
  review,'' \emph{Turkish Journal of Computer and Mathematics Education},
  vol.~12, no.~4, pp. 1041--1053, 2021.

\bibitem{kaur2021systematic}
N.~Kaur, A.~Kumar, and R.~Kumar, ``A systematic review on task scheduling in
  fog computing: Taxonomy, tools, challenges, and future directions,''
  \emph{Concurrency and Computation: Practice and Experience}, vol.~33, no.~21,
  p. e6432, 2021.

\bibitem{zhang2021joint}
J.~Zhang, X.~Zhou, T.~Ge, X.~Wang, and T.~Hwang, ``Joint task scheduling and
  containerizing for efficient edge computing,'' \emph{IEEE Transactions on
  Parallel and Distributed Systems}, vol.~32, no.~8, pp. 2086--2100, 2021.

\bibitem{munjal2022big}
G.~Munjal and M.~Kumar, ``Big data: Related technologies and applications,'' in
  \emph{Transforming Management with AI, Big-Data, and IoT}.\hskip 1em plus
  0.5em minus 0.4em\relax Springer, 2022, pp. 85--98.

\bibitem{wang2021privacy}
T.~Wang, Q.~Yang, X.~Shen, T.~R. Gadekallu, W.~Wang, and K.~Dev, ``A
  privacy-enhanced retrieval technology for the cloud-assisted internet of
  things,'' \emph{IEEE Transactions on Industrial Informatics}, vol.~18, no.~7,
  pp. 4981--4989, 2021.

\bibitem{awaysheh2021big}
F.~M. Awaysheh, M.~Alazab, S.~Garg, D.~Niyato, and C.~Verikoukis, ``Big data
  resource management \& networks: Taxonomy, survey, and future directions,''
  \emph{IEEE Communications Surveys \& Tutorials}, 2021.

\bibitem{tao2018digital}
F.~Tao, H.~Zhang, A.~Liu, and A.~Y. Nee, ``Digital twin in industry:
  State-of-the-art,'' \emph{IEEE Transactions on Industrial Informatics},
  vol.~15, no.~4, pp. 2405--2415, 2018.

\bibitem{liu2021review}
M.~Liu, S.~Fang, H.~Dong, and C.~Xu, ``Review of digital twin about concepts,
  technologies, and industrial applications,'' \emph{Journal of Manufacturing
  Systems}, vol.~58, pp. 346--361, 2021.

\bibitem{fan2021disaster}
C.~Fan, C.~Zhang, A.~Yahja, and A.~Mostafavi, ``Disaster city digital twin: A
  vision for integrating artificial and human intelligence for disaster
  management,'' \emph{International Journal of Information Management},
  vol.~56, p. 102049, 2021.

\bibitem{watson2022virtual}
R.~Watson \emph{et~al.}, ``The virtual economy of the metaverse: Computer
  vision and deep learning algorithms, customer engagement tools, and
  behavioral predictive analytics,'' \emph{Linguistic and Philosophical
  Investigations}, no.~21, pp. 41--56, 2022.

\bibitem{lee2019decentralized}
J.~Y. Lee, ``A decentralized token economy: How blockchain and cryptocurrency
  can revolutionize business,'' \emph{Business Horizons}, vol.~62, no.~6, pp.
  773--784, 2019.

\bibitem{lee2021creators}
L.-H. Lee, Z.~Lin, R.~Hu, Z.~Gong, A.~Kumar, T.~Li, S.~Li, and P.~Hui, ``When
  creators meet the metaverse: A survey on computational arts,'' \emph{arXiv
  preprint arXiv:2111.13486}, 2021.

\bibitem{kye2021educational}
B.~Kye, N.~Han, E.~Kim, Y.~Park, and S.~Jo, ``Educational applications of
  metaverse: possibilities and limitations,'' \emph{Journal of Educational
  Evaluation for Health Professions}, vol.~18, 2021.

\bibitem{mystakidis2022Metaverse}
S.~Mystakidis, ``Metaverse,'' \emph{Encyclopedia}, vol.~2, no.~1, pp. 486--497,
  2022.

\bibitem{hirsh2022whole}
K.~Hirsh-Pasek, J.~Zosh, H.~S. Hadani, R.~M. Golinkoff, K.~Clark, C.~Donohue,
  and E.~Wartella, ``A whole new world: Education meets the metaverse,''
  \emph{Policy}, 2022.

\bibitem{almarzouqi2022prediction}
A.~Almarzouqi, A.~Aburayya, and S.~A. Salloum, ``Prediction of user's intention
  to use metaverse system in medical education: A hybrid {SEM-ML} learning
  approach,'' \emph{IEEE Access}, vol.~10, pp. 43\,421--43\,434, 2022.

\bibitem{jeong2022innovative}
H.~Jeong, Y.~Yi, and D.~Kim, ``An innovative e-commerce platform incorporating
  metaverse to live commerce,'' \emph{International Journal of Innovative
  Computing, Information and Control}, vol.~18, no.~1, pp. 221--229, 2022.

\bibitem{gadalla2013Metaverse}
E.~Gadalla, K.~Keeling, and I.~Abosag, ``Metaverse-retail service quality: A
  future framework for retail service quality in the {3D} internet,''
  \emph{Journal of Marketing Management}, vol.~29, no. 13-14, pp. 1493--1517,
  2013.

\bibitem{werner2022use}
H.~Werner, G.~Ribeiro, V.~Arcoverde, J.~Lopes, and L.~Velho, ``The use of
  metaverse in fetal medicine and gynecology,'' \emph{European Journal of
  Radiology}, vol. 150, 2022.

\bibitem{mozumder2022overview}
M.~A.~I. Mozumder, M.~M. Sheeraz, A.~Athar, S.~Aich, and H.-C. Kim, ``Overview:
  Technology roadmap of the future trend of metaverse based on {IoT},
  blockchain, {AI} technique, and medical domain metaverse activity,'' in
  \emph{24th International Conference on Advanced Communication
  Technology}.\hskip 1em plus 0.5em minus 0.4em\relax IEEE, 2022, pp. 256--261.

\bibitem{chengoden2022Metaverse}
R.~Chengoden, N.~Victor, T.~Huynh-The, G.~Yenduri, R.~H. Jhaveri, M.~Alazab,
  S.~Bhattacharya, P.~Hegde, P.~K.~R. Maddikunta, and T.~R. Gadekallu,
  ``Metaverse for healthcare: A survey on potential applications, challenges
  and future directions,'' \emph{arXiv preprint arXiv:2209.04160}, 2022.

\bibitem{skalidis2022cardioverse}
I.~Skalidis, O.~Muller, and S.~Fournier, ``Cardioverse: The cardiovascular
  medicine in the era of metaverse,'' \emph{Trends in Cardiovascular Medicine},
  2022.

\bibitem{locurcio2022dental}
L.~Locurcio, ``Dental education in the metaverse,'' \emph{British Dental
  Journal}, vol. 232, no.~4, pp. 191--191, 2022.

\bibitem{lee2011self}
H.~Lee and A.~Banerjee, ``A self-configurable large-scale virtual manufacturing
  environment for collaborative designers,'' \emph{Virtual Reality}, vol.~15,
  no.~1, pp. 21--40, 2011.

\bibitem{lin2022towards}
Z.~Lin, P.~Xiangli, Z.~Li, F.~Liang, and A.~Li, ``Towards metaverse
  manufacturing: A blockchain-based trusted collaborative governance system,''
  in \emph{The 4th International Conference on Blockchain Technology}, 2022,
  pp. 171--177.

\bibitem{choi2022working}
H.-Y. Choi, ``Working in the metaverse: Does telework in a metaverse office
  have the potential to reduce population pressure in megacities? evidence from
  young adults in seoul, south korea,'' \emph{Sustainability}, vol.~14, no.~6,
  p. 3629, 2022.

\bibitem{lawrence2021project}
J.~Lawrence, D.~B. Goldman, S.~Achar, G.~M. Blascovich, J.~G. Desloge,
  T.~Fortes, E.~M. Gomez, S.~H{\"a}berling, H.~Hoppe, A.~Huibers \emph{et~al.},
  ``Project starline: A high-fidelity telepresence system,'' 2021.

\bibitem{gupta2021fusion}
R.~Gupta, A.~Kumari, and S.~Tanwar, ``Fusion of blockchain and artificial
  intelligence for secure drone networking underlying 5g communications,''
  \emph{Transactions on Emerging Telecommunications Technologies}, vol.~32,
  no.~1, p. e4176, 2021.

\bibitem{ni2019research}
Y.~Ni, J.~Liang, X.~Shi, and D.~Ban, ``Research on key technology in 5g mobile
  communication network,'' in \emph{International Conference on Intelligent
  Transportation, Big Data \& Smart City}.\hskip 1em plus 0.5em minus
  0.4em\relax IEEE, 2019, pp. 199--201.

\bibitem{cheng2022will}
R.~Cheng, N.~Wu, S.~Chen, and B.~Han, ``Will metaverse be nextg internet?
  vision, hype, and reality,'' \emph{arXiv preprint arXiv:2201.12894}, 2022.

\bibitem{chang2008bigtable}
F.~Chang, J.~Dean, S.~Ghemawat, W.~C. Hsieh, D.~A. Wallach, M.~Burrows,
  T.~Chandra, A.~Fikes, and R.~E. Gruber, ``Bigtable: A distributed storage
  system for structured data,'' \emph{ACM Transactions on Computer Systems},
  vol.~26, no.~2, pp. 1--26, 2008.

\bibitem{al2010case}
S.~Al-Kiswany, A.~Gharaibeh, and M.~Ripeanu, ``The case for a versatile storage
  system,'' \emph{ACM SIGOPS Operating Systems Review}, vol.~44, no.~1, pp.
  10--14, 2010.

\bibitem{tang2022roadmap}
F.~Tang, X.~Chen, M.~Zhao, and N.~Kato, ``The roadmap of communication and
  networking in 6g for the metaverse,'' \emph{IEEE Wireless Communications},
  2022.

\bibitem{tang2021computing}
X.~Tang, C.~Cao, Y.~Wang, S.~Zhang, Y.~Liu, M.~Li, and T.~He, ``Computing power
  network: The architecture of convergence of computing and networking towards
  6g requirement,'' \emph{China Communications}, vol.~18, no.~2, pp. 175--185,
  2021.

\bibitem{lim2022realizing}
W.~Y.~B. Lim, Z.~Xiong, D.~Niyato, X.~Cao, C.~Miao, S.~Sun, and Q.~Yang,
  ``Realizing the metaverse with edge intelligence: A match made in heaven,''
  \emph{arXiv preprint arXiv:2201.01634}, 2022.

\bibitem{nawaratne2018self}
R.~Nawaratne, D.~Alahakoon, D.~De~Silva, P.~Chhetri, and N.~Chilamkurti,
  ``Self-evolving intelligent algorithms for facilitating data interoperability
  in {IoT} environments,'' \emph{Future Generation Computer Systems}, vol.~86,
  pp. 421--432, 2018.

\bibitem{gonzalez2019interoperability}
R.~Gonzalez-Usach, D.~Yacchirema, M.~Julian, and C.~E. Palau,
  ``Interoperability in {IoT},'' in \emph{Handbook of Research on Big Data and
  the IoT}.\hskip 1em plus 0.5em minus 0.4em\relax IGI Global, 2019, pp.
  149--173.

\bibitem{bhattacharya2021coalition}
P.~Bhattacharya, D.~Saraswat, A.~Dave, M.~Acharya, S.~Tanwar, G.~Sharma, and
  I.~E. Davidson, ``Coalition of 6g and blockchain in {AR/VR} space: Challenges
  and future directions,'' \emph{IEEE Access}, vol.~9, pp. 168\,455--168\,484,
  2021.

\bibitem{dong2015Secure}
X.~Dong, R.~Li, H.~He, W.~Zhou, Z.~Xue, and H.~Wu, ``Secure sensitive data
  sharing on a big data platform,'' \emph{Tsinghua Science and Technology},
  vol.~20, no.~1, pp. 72--80, 2015.

\bibitem{chen2019fade}
Y.~Chen, J.~Guo, C.~Li, and W.~Ren, ``{FaDe}: a blockchain-based fair data
  exchange scheme for big data sharing,'' \emph{Future Internet}, vol.~11,
  no.~11, p. 225, 2019.

\bibitem{yu2021blockchain}
K.~Yu, L.~Tan, M.~Aloqaily, H.~Yang, and Y.~Jararweh, ``Blockchain-enhanced
  data sharing with traceable and direct revocation in {IIoT},'' \emph{IEEE
  Transactions on Industrial Informatics}, vol.~17, no.~11, pp. 7669--7678,
  2021.

\bibitem{egliston2021critical}
B.~Egliston and M.~Carter, ``Critical questions for facebook’s virtual
  reality: data, power and the metaverse,'' \emph{Internet Policy Review},
  vol.~10, no.~4, 2021.

\bibitem{von2013information}
R.~Von~Solms and J.~Van~Niekerk, ``From information security to cyber
  security,'' \emph{Computers \& Security}, vol.~38, pp. 97--102, 2013.

\bibitem{kyto2017augmenting}
M.~Kyt{\"o} and D.~McGookin, ``Augmenting multi-party face-to-face interactions
  amongst strangers with user generated content,'' \emph{Computer Supported
  Cooperative Work}, vol.~26, no.~4, pp. 527--562, 2017.

\bibitem{carmigniani2011augmented}
J.~Carmigniani, B.~Furht, M.~Anisetti, P.~Ceravolo, E.~Damiani, and M.~Ivkovic,
  ``Augmented reality technologies, systems and applications,''
  \emph{Multimedia Tools and Applications}, vol.~51, no.~1, pp. 341--377, 2011.

\bibitem{shah2014survey}
K.~N. Shah, K.~R. Rathod, and S.~J. Agravat, ``A survey on human computer
  interaction mechanism using finger tracking,'' \emph{arXiv preprint
  arXiv:1402.0693}, 2014.

\bibitem{clay2019eye}
V.~Clay, P.~K{\"o}nig, and S.~Koenig, ``Eye tracking in virtual reality,''
  \emph{Journal of Eye Movement Research}, vol.~12, no.~1, 2019.

\bibitem{conti2016survey}
M.~Conti, N.~Dragoni, and V.~Lesyk, ``A survey of man in the middle attacks,''
  \emph{IEEE Communications Surveys \& Tutorials}, vol.~18, no.~3, pp.
  2027--2051, 2016.

\bibitem{bhushan2017man}
B.~Bhushan, G.~Sahoo, and A.~K. Rai, ``Man-in-the-middle attack in wireless and
  computer networking—a review,'' in \emph{3rd International Conference on
  Advances in Computing, Communication \& Automation}.\hskip 1em plus 0.5em
  minus 0.4em\relax IEEE, 2017, pp. 1--6.

\bibitem{masdari2016survey}
M.~Masdari and M.~Jalali, ``A survey and taxonomy of {DoS} attacks in cloud
  computing,'' \emph{Security and Communication Networks}, vol.~9, no.~16, pp.
  3724--3751, 2016.

\bibitem{zhijun2020low}
W.~Zhijun, L.~Wenjing, L.~Liang, and Y.~Meng, ``Low-rate {DoS} attacks,
  detection, defense, and challenges: A survey,'' \emph{IEEE Access}, vol.~8,
  pp. 43\,920--43\,943, 2020.

\bibitem{greenberg1998mobile}
M.~S. Greenberg, J.~C. Byington, and D.~G. Harper, ``Mobile agents and
  security,'' \emph{IEEE Communications Magazine}, vol.~36, no.~7, pp. 76--85,
  1998.

\bibitem{sengupta2020comprehensive}
J.~Sengupta, S.~Ruj, and S.~D. Bit, ``A comprehensive survey on attacks,
  security issues and blockchain solutions for {IoT and IIoT},'' \emph{Journal
  of Network and Computer Applications}, vol. 149, p. 102481, 2020.

\bibitem{dillon2010cloud}
T.~Dillon, C.~Wu, and E.~Chang, ``Cloud computing: issues and challenges,'' in
  \emph{24th IEEE International Conference on Advanced Information Networking
  and Applications}.\hskip 1em plus 0.5em minus 0.4em\relax Ieee, 2010, pp.
  27--33.

\bibitem{qian2009cloud}
L.~Qian, Z.~Luo, Y.~Du, and L.~Guo, ``Cloud computing: An overview,'' in
  \emph{IEEE International Conference on Cloud Computing}.\hskip 1em plus 0.5em
  minus 0.4em\relax Springer, 2009, pp. 626--631.

\bibitem{chen2012data}
D.~Chen and H.~Zhao, ``Data security and privacy protection issues in cloud
  computing,'' in \emph{International Conference on Computer Science and
  Electronics Engineering}, vol.~1.\hskip 1em plus 0.5em minus 0.4em\relax
  IEEE, 2012, pp. 647--651.

\bibitem{jian2021hybrid}
M.-S. Jian and J.~M.-T. Wu, ``Hybrid internet of things {(IoT)} data
  transmission security corresponding to device verification,'' \emph{Journal
  of Ambient Intelligence and Humanized Computing}, pp. 1--10, 2021.

\bibitem{jain2016big}
P.~Jain, M.~Gyanchandani, and N.~Khare, ``Big data privacy: a technological
  perspective and review,'' \emph{Journal of Big Data}, vol.~3, no.~1, pp.
  1--25, 2016.

\bibitem{uzuner2007evaluating}
{\"O}.~Uzuner, Y.~Luo, and P.~Szolovits, ``Evaluating the state-of-the-art in
  automatic de-identification,'' \emph{Journal of the American Medical
  Informatics Association}, vol.~14, no.~5, pp. 550--563, 2007.

\bibitem{sweeney2002k}
L.~Sweeney, ``k-anonymity: A model for protecting privacy,''
  \emph{International Journal of Uncertainty, Fuzziness and Knowledge-Based
  Systems}, vol.~10, no.~05, pp. 557--570, 2002.

\bibitem{liu2019voting}
Y.~Liu and Q.~Zhao, ``E-voting scheme using secret sharing and k-anonymity,''
  \emph{World Wide Web}, vol.~22, no.~4, pp. 1657--1667, 2019.

\bibitem{machanavajjhala2007diversity}
A.~Machanavajjhala, D.~Kifer, J.~Gehrke, and M.~Venkitasubramaniam,
  ``l-diversity: Privacy beyond k-anonymity,'' \emph{ACM Transactions on
  Knowledge Discovery from Data}, vol.~1, no.~1, pp. 3--es, 2007.

\bibitem{mehta2019improved}
B.~B. Mehta and U.~P. Rao, ``Improved l-diversity: Scalable anonymization
  approach for privacy preserving big data publishing,'' \emph{Journal of King
  Saud University-Computer and Information Sciences}, 2019.

\bibitem{friedman2010data}
A.~Friedman and A.~Schuster, ``Data mining with differential privacy,'' in
  \emph{The 16th ACM SIGKDD International Conference on Knowledge Discovery and
  Data Mining}, 2010, pp. 493--502.

\bibitem{guo2021practical}
J.~Guo, M.~Yang, and B.~Wan, ``A practical privacy-preserving publishing
  mechanism based on personalized k-anonymity and temporal differential privacy
  for wearable iot applications,'' \emph{Symmetry}, vol.~13, no.~6, p. 1043,
  2021.

\bibitem{acar2018survey}
A.~Acar, H.~Aksu, A.~S. Uluagac, and M.~Conti, ``A survey on homomorphic
  encryption schemes: Theory and implementation,'' \emph{ACM Computing
  Surveys}, vol.~51, no.~4, pp. 1--35, 2018.

\bibitem{karray2008human}
F.~Karray, M.~Alemzadeh, J.~Abou~Saleh, and M.~N. Arab, ``Human-computer
  interaction: Overview on state of the art,'' \emph{International Journal on
  Smart Sensing and Intelligent Systems}, vol.~1, no.~1, p. 137, 2008.

\bibitem{zhu2022metaaid}
H.~Zhu, ``Metaaid: A flexible framework for developing metaverse applications
  via ai technology and human editing,'' \emph{arXiv preprint
  arXiv:2204.01614}, 2022.

\bibitem{zhao2020expression}
G.~Zhao, H.~Yang, and M.~Yu, ``Expression recognition method based on a
  lightweight convolutional neural network,'' \emph{IEEE Access}, vol.~8, pp.
  38\,528--38\,537, 2020.

\bibitem{santavas2020attention}
N.~Santavas, I.~Kansizoglou, L.~Bampis, E.~Karakasis, and A.~Gasteratos,
  ``Attention! a lightweight 2d hand pose estimation approach,'' \emph{IEEE
  Sensors Journal}, vol.~21, no.~10, pp. 11\,488--11\,496, 2020.

\bibitem{aftab2022light}
A.~Aftab, A.~Morsali, S.~Ghaemmaghami, and B.~Champagne, ``Light-sernet: A
  lightweight fully convolutional neural network for speech emotion
  recognition,'' in \emph{IEEE International Conference on Acoustics, Speech
  and Signal Processing}.\hskip 1em plus 0.5em minus 0.4em\relax IEEE, 2022,
  pp. 6912--6916.

\bibitem{ma2019lightweight}
S.~Ma, T.~Zhang, A.~Wu, and X.~Zhao, ``Lightweight and privacy-preserving data
  aggregation for mobile multimedia security,'' \emph{IEEE Access}, vol.~7, pp.
  114\,131--114\,140, 2019.

\bibitem{deebak2021lightweight}
B.~D. Deebak and A.-T. Fadi, ``Lightweight authentication for {IoT}/cloud-based
  forensics in intelligent data computing,'' \emph{Future Generation Computer
  Systems}, vol. 116, pp. 406--425, 2021.

\bibitem{stojanovski2020city}
T.~Stojanovski, J.~Partanen, I.~Samuels, P.~Sanders, and C.~Peters, ``City
  information modelling ({CIM}) and digitizing urban design practices,''
  \emph{Built Environment}, vol.~46, no.~4, pp. 637--646, 2020.

\bibitem{xue2021semantic}
F.~Xue, L.~Wu, and W.~Lu, ``Semantic enrichment of building and city
  information models: A ten-year review,'' \emph{Advanced Engineering
  Informatics}, vol.~47, p. 101245, 2021.

\bibitem{fu2020bim}
M.~Fu and R.~Liu, ``{BIM}-based automated determination of exit sign direction
  for intelligent building sign systems,'' \emph{Automation in Construction},
  vol. 120, p. 103353, 2020.

\bibitem{barassi2012does}
V.~Barassi and E.~Trer{\'e}, ``Does {Web} 3.0 come after {Web} 2.0?
  deconstructing theoretical assumptions through practice,'' \emph{New Media \&
  Society}, vol.~14, no.~8, pp. 1269--1285, 2012.

\bibitem{rudman2016defining}
R.~Rudman and R.~Bruwer, ``Defining {Web} 3.0: opportunities and challenges,''
  \emph{The Electronic Library}, 2016.

\bibitem{kshetri2022web}
N.~Kshetri, ``{Web} 3.0 and the metaverse shaping organizations’ brand and
  product strategies,'' \emph{IT Professional}, vol.~24, no.~02, pp. 11--15,
  2022.

\bibitem{murugesan2007understanding}
S.~Murugesan, ``Understanding {Web} 2.0,'' \emph{IT Professional}, vol.~9,
  no.~4, pp. 34--41, 2007.

\bibitem{o2007web}
T.~O'reilly, ``What is {Web} 2.0: Design patterns and business models for the
  next generation of software,'' \emph{Communications \& Strategies}, no.~1,
  p.~17, 2007.

\end{thebibliography}

\end{document}